\documentclass[ aps, twocolumn, nofootinbib, superscriptaddress ]{revtex4-2}

\usepackage{graphicx}  
\usepackage{dcolumn}   
\usepackage{bm}        
\usepackage{amssymb,amsfonts,amsmath,physics}   
\usepackage{cancel}
\usepackage{mathtools}
\usepackage{slashed}
\usepackage{float}
\usepackage[dvipsnames]{xcolor}

\usepackage{tikz-feynman}
\tikzfeynmanset{compat=1.1.0}
\usetikzlibrary{arrows}
\tikzset{	
	vertex/.style={circle,draw, minimum size=1.5em},	
	edge/.style={->,> = latex'}	
}
\usepackage[bookmarks, breaklinks, colorlinks,urlcolor=blue, citecolor=red, 
linkcolor=blue]{hyperref}
\usepackage[normalem]{ulem}

\hypersetup{
	colorlinks=true,
	linkcolor=Red,
	filecolor=magenta,
	urlcolor=Blue,
	citecolor=PineGreen}

\newcommand{\be}{\begin{eqnarray*}}
	\newcommand{\ee}{\end{eqnarray*}}

\newcommand{\bee}{\begin{eqnarray}}
	\newcommand{\eee}{\end{eqnarray}}
\newcommand{\beeq}{\begin{equation}}
	\newcommand{\eeq}{\end{equation}}

\usepackage{xspace}

\newcommand{\ba}{\begin{array}}
	\newcommand{\ea}{\end{array}}
\newcommand{\bd}{\begin{displaymath}}
	\newcommand{\ed}{\end{displaymath}}
\newcommand{\besub}{\begin{subequations}}
	\newcommand{\eesub}{\end{subequations}}
\newcommand{\bea}{\begin{eqnarray}}
	\newcommand{\eea}{\end{eqnarray}}

\newcommand{\sla}[1]{/\!\!\!#1}

\def\q2 {q^2}

\usepackage{tikz}
\usetikzlibrary{arrows,shapes}
\usetikzlibrary{trees}
\usetikzlibrary{matrix,arrows} 				
\usetikzlibrary{positioning}				
\usetikzlibrary{calc,through}				
\usetikzlibrary{decorations.pathreplacing}  
\usepackage{pgffor}							

\usetikzlibrary{decorations.pathmorphing}	
\usetikzlibrary{decorations.markings}
\tikzset{
	vector/.style={decorate, decoration={snake}, draw},
	provector/.style={decorate, decoration={snake,amplitude=2.5pt}, draw},
	antivector/.style={decorate, decoration={snake,amplitude=-2.5pt}, draw},
	fermion/.style={draw=black, postaction={decorate},
		decoration={markings,mark=at position .55 with {\arrow[draw=black]{>}}}},
	fermionbar/.style={draw=black, postaction={decorate},
		decoration={markings,mark=at position .55 with {\arrow[draw=black]{<}}}},
	fermionnoarrow/.style={draw=black},
	gluon/.style={decorate, draw=black,
		decoration={coil,amplitude=4pt, segment length=5pt}},
	scalar/.style={dashed,draw=black, postaction={decorate},
		decoration={markings,mark=at position .55 with {\arrow[draw=black]{>}}}},
	scalarbar/.style={dashed,draw=black, postaction={decorate},
		decoration={markings,mark=at position .55 with {\arrow[draw=black]{<}}}},
	scalarnoarrow/.style={dashed,draw=black},
	electron/.style={draw=black, postaction={decorate},
		decoration={markings,mark=at position .55 with {\arrow[draw=black]{>}}}},
	bigvector/.style={decorate, decoration={snake,amplitude=4pt}, draw},
}

\tikzstyle{block} = [draw, rectangle, 
minimum height=3em, minimum width=6em]

\begin{document}

    \title{Leptogenesis with a GeV-Scale Quasi-Dirac Neutrino}

	\author{Stephen F. King}
	\email{S.F.king@soton.ac.uk}
	\affiliation{School of Physics and Astronomy, University of Southampton,
		Southampton SO17 1BJ, United Kingdom}
	
	\author{Soumen Kumar Manna}
	\email{skmanna2021@gmail.com}
	\affiliation{Department of Physics, Indian Institute of Technology Bombay, Mumbai 400076, India}
	
	\author{Rishav Roshan}
	\email{rishav.roshan@gmail.com}
	\affiliation{School of Physics and Astronomy, University of Southampton,
		Southampton SO17 1BJ, United Kingdom}
	
	\author{Arunansu Sil}
	\email{asil@iitg.ac.in}
	\affiliation{Department of Physics, Indian Institute of Technology Guwahati, Assam-781039, India}
	
	\begin{abstract} 
We discuss leptogenesis in a model with a GeV-scale quasi-Dirac neutral lepton, comprising mainly sterile states, which is accessible to future experiments.
In particular, we show that Higgs-decay leptogenesis can naturally arise from the type-I seesaw mechanism with spontaneously broken global lepton number together with a discrete $Z_2$ symmetry, yielding two quasi-degenerate right-handed neutrinos (RHNs) which, to excellent approximation, form a single Dirac neutrino, with a tiny mass splitting as required for resonant enhancement generated by higher-dimensional lepton-number-violating operators. Incorporating finite-temperature effects, we show that the observed baryon asymmetry can be successfully generated through thermal Higgs decays for 0.1-10 GeV mass quasi-Dirac neutrinos, with the CP asymmetry arising from the interplay between the vacuum and thermal contributions to the mass splitting. In the parameter region of successful leptogenesis, the quasi-Dirac neutrino is accessible to future heavy neutral lepton searches at SHiP, MATHUSLA and FCC-ee. In addition, the same symmetry-breaking sector gives rise to unstable domain walls whose annihilation can generate a stochastic gravitational-wave background, providing a complementary probe of the framework.

	\end{abstract}
	\maketitle


\section{Introduction}
\label{sec.introduction}

The origin of the observed matter-antimatter asymmetry of the Universe remains one of the strongest motivations for physics beyond the Standard Model (SM). Among the proposed explanations, leptogenesis ~\cite{Fukugita:1986hr,Buchmuller:2005eh,Davidson:2008bu} provides one of the most elegant and well-motivated mechanisms, as it is intimately connected to another outstanding puzzle in particle physics, the origin of neutrino masses. The simplest realization of this framework extends the SM by introducing right-handed neutrinos (RHNs), which are singlets under the SM gauge group and can possess lepton-number-violating Majorana masses. These heavy states naturally generate tiny active neutrino masses through the seesaw mechanism \cite{Minkowski:1977sc,Mohapatra:1979ia,Schechter:1980gr}, while their out of equilibrium, CP-violating decays in the early Universe produce a lepton asymmetry that is subsequently converted into the observed baryon asymmetry through electroweak sphaleron processes. Thus, the seesaw mechanism and leptogenesis provide a unified framework for simultaneously explaining neutrino masses and the baryon asymmetry of the Universe.

In the conventional thermal leptogenesis scenario, successful baryogenesis typically requires the lightest RHN to be very heavy, with masses above $\sim 10^9$ GeV \cite{Davidson:2002qv} for a hierarchical RHN spectrum. Although such high masses naturally accommodate the observed baryon asymmetry, they render the RHNs essentially inaccessible to direct experimental searches, making the mechanism difficult to test. This limitation has motivated extensive studies of low-scale leptogenesis scenarios, in which the RHN masses can lie at or below the TeV scale, making them attractive from experimental perspective. Such scenarios rely on mechanisms capable of enhancing the CP asymmetry, the most prominent example being resonant leptogenesis\cite{Pilaftsis:1997jf,Pilaftsis:2003gt,Ibarra:2003up,BhupalDev:2014pfm,BhupalDev:2014oar}. Heavy neutral leptons \cite{Abdullahi:2022jlv} with masses below the electroweak scale are actively being searched for at LHC experiments (ATLAS \cite{ATLAS:2012yve, ATLAS:2025uah}, CMS \cite{CMS:2012qbp, CMS:2024xdq}, and LHCb \cite{LHCb:2020wxx,LHCb:2025ymr}), while lighter mass regions are explored by intensity-frontier experiments such as NA62 \cite{Drewes:2018gkc,NA62:2020mcv} and FASER \cite{Feng:2017uoz, FASER:2022hcn}. Future facilities, including SHiP \cite{SHiP:2018xqw}, MATHUSLA \cite{Curtin:2018mvb} and proposed lepton colliders such as FCC-ee \cite{FCC:2018evy}, are expected to significantly extend the discovery reach for GeV-scale RHNs.

Although lowering the RHN mass scale substantially improves the prospects for experimental verification, it also makes successful leptogenesis considerably more challenging. A well-established solution is resonant leptogenesis, in which the CP asymmetry generated in RHN decays is resonantly enhanced through the self-energy contribution when at least two RHNs are nearly degenerate in mass. This enhancement makes it possible to generate the observed baryon asymmetry even for RHN masses around the electroweak or TeV scale. However, the success of this mechanism typically relies on an exceptionally small RHN mass splitting, often together with suppressed Yukawa couplings, to ensure the out-of-equilibrium condition. Although phenomenologically successful, the required near-degeneracy is generally introduced as an additional assumption, whose theoretical origin remains unclear.

In our earlier work~\cite{King:2024idj}, we proposed a symmetry-based framework that naturally realizes the quasi-degenerate RHN spectrum
required for resonant leptogenesis, with RHNs in the mass range $\mathcal{O}(10^6) - \mathcal{O}(10^{13})$ GeV.
The model extends the SM by a global lepton-number symmetry, $U(1)_L$, together with a discrete $Z_2$ symmetry, and includes two RHNs as well as a scalar responsible for the spontaneous breaking of $U(1)_L$. In the symmetry limit, the spontaneous breaking of $U(1)_L$ gives rise to a heavy Dirac neutrino, comprised of two exactly degenerate RHNs, through the renormalizable interactions. A tiny mass splitting, essential for resonant enhancement, is subsequently generated by dimension-five operators that explicitly break the global lepton-number symmetry while preserving the $Z_2$ symmetry. Consequently, the small RHN mass splitting emerges naturally from the underlying symmetry structure, providing a theoretically well-motivated origin for one of the key ingredients of resonant leptogenesis.

In the present paper, motivated by experimental testability,
we consider a similar model to that previously considered~\cite{King:2024idj}, but with a much lower 
GeV-scale quasi-Dirac neutral lepton, comprising the two closely degenerate RHN states. Although the construction described above provides a natural origin for the required tiny RHN mass splitting, pushing the RHN mass scale below the electroweak scale presents an additional challenge. The generated lepton asymmetry must be converted into a baryon asymmetry through electroweak sphaleron processes, which become ineffective below the sphaleron decoupling temperature, $T_{\rm sph}\simeq131.7$ GeV \cite{DOnofrio:2014rug}. 
An elegant way to overcome this limitation is provided by Higgs-decay leptogenesis \cite{Hambye:2016sby,Hambye:2017elz}. Owing to finite-temperature effects, the Higgs boson acquires a thermal mass in the early Universe, rendering the decay of it into RHNs and lepton doublets ($H\rightarrow L N$) kinematically allowed before electroweak symmetry breaking. The resulting CP-violating Higgs decays can generate the required lepton asymmetry while electroweak sphaleron processes are still active, thereby extending the possibility of successful leptogenesis to RHNs with masses well below the electroweak scale. Consequently, successful leptogenesis with sub-electroweak-scale RHNs requires the asymmetry to be generated within the relatively narrow temperature interval between the electroweak phase transition $T_{\rm EW} \sim 160$ GeV and sphaleron freeze-out. 
Since Higgs-decay leptogenesis likewise relies on resonantly enhanced CP asymmetry, it also requires a tiny mass splitting between the RHNs. Our symmetry-based framework therefore provides a natural setting for realizing this mechanism. 

Within the above framework, we investigate Higgs-decay leptogenesis and demonstrate that the observed baryon asymmetry can be successfully generated for a GeV-scale quasi-Dirac neutrino, comprising two closely degenerate 
RHNs, through a purely thermal CP asymmetry arising from the interplay between the symmetry-induced vacuum mass splitting and finite-temperature effects. We further identify the experimentally viable parameter space that can be probed in future heavy neutral lepton searches, particularly at facilities such as SHiP and FCC-ee. As a complementary consequence of the framework, we also briefly discuss the stochastic gravitational-wave (GW) signals associated with domain-wall (DW) annihilation \cite{Saikawa:2017hiv,Dunsky:2021tih,Sakharov:2021dim,Ferreira:2022zzo, An:2023idh,  Bai:2023cqj,King:2023ayw,Roshan:2024qnv,Roshan:2026xpf} induced by the same symmetry-breaking sector.

We also briefly mention other alternative leptogenesis mechanisms, relevant for the GeV-scale RHN mass regime, for example the Akhmedov-Rubakov-Smirnov (ARS) mechanism \cite{Akhmedov:1998qx}, where the baryon asymmetry originates from CP-violating oscillations of nearly degenerate RHNs together with flavour-dependent washout effects. Another distinct possibility is resonant leptogenesis through RHN decays in the presence of a strongly first-order electroweak phase transition \cite{Bhandari:2025ufp}, where the effective sphaleron decoupling temperature is lowered below its SM value, allowing successful baryogenesis with sub-electroweak-scale RHNs. In contrast, the present framework realizes low-scale leptogenesis through CP-violating thermal Higgs decays, with the tiny RHN mass splitting required for resonant enhancement emerging naturally from the underlying symmetry structure. Together, these complementary mechanisms highlight the rich landscape of experimentally testable low-scale leptogenesis with a GeV-scale quasi-Dirac neutrino.

This paper is organized as follows. In Section~\ref{sec.model}, we introduce our model, which involves two RHNs and the lepton-number-symmetry-breaking scalar field $\Phi$, featuring both global $U(1)_L$-conserving and -breaking terms. In Section~\ref{sec.leptogenesis}, we discuss low-scale leptogenesis from Higgs decay. In Section~\ref{ARS}, we compare the Higgs decay mechanism with the ARS mechanism of asymmetry generation, followed by the results and analysis of our work in Section~\ref{results}. Next, in Section~\ref{sec:scan}, we present the region of parameter space that can be probed in various experiments. In Section~\ref{majoron}, we briefly address the fate of majorons in our model. In Section~\ref{sec:GW}, we then discuss the GW production resulting from DW annihilation. Finally, we conclude in Section~\ref{conclusions}.


	\section{SET-UP}
	\label{sec.model}
Building upon our earlier work~\cite{King:2024idj}, we consider here a minimal type-I seesaw framework with two RHNs \footnote{The cosmological consequences of the third RHN are explored in Ref.~\cite{Datta:2021elq,Roshan:2025mwl,King:2026pou}.} $\mathcal{N}_i$ and a complex SM gauge singlet scalar field $\Phi$. The SM gauge symmetry is extended by a global lepton number symmetry $U(1)_L$, supplemented by a discrete $Z_2$ symmetry. Both symmetries are spontaneously broken when $\Phi$ acquires a non-zero vacuum expectation value (VEV). The charge assignments for the beyond the SM (BSM) fields under the $U(1)_L$ 
and $Z_2$ symmetries, are summarized in Table~\ref{tab-charge}.
\begin{table}[h]
		\begin{center}
			\vskip 0.5 cm
			\begin{tabular}{|c|c|c|c|}
				\hline
				Symmetries	&	$\Phi$ & $\mathcal{N}_1$ &  $\mathcal{N}_2$   \\
				\hline
				$U(1)_L$       &  $-2$     &  1 & 1  \\                 
				\hline
				$Z_2$    & $-$ &+ & $-$  \\                 
				\hline
			\end{tabular}
		\end{center}  
		\caption{Beyond the SM particles and their charges under global $U(1)_L$ and $Z_2$ symmetries.}
	\label{tab-charge}
\end{table}

We now write down the Lagrangian, involving the BSM fields, invariant under these symmetries as 
\beeq
	-\mathcal{L}_{\rm c}\supset  \frac{f}{2}\Phi \overline{\mathcal{N}_1^C}\mathcal{N}_2+y_{\alpha 1}\overline{L_\alpha} \tilde{H}\mathcal{N}_1 +h.c.,
	\label{eq:seesaw}
\eeq
where $H$ and $L$ denote the SM Higgs doublet and lepton doublet respectively. While the symmetries permit the Yukawa coupling $y_{\alpha 1}$ of $\mathcal{N}_1$ to the SM lepton and Higgs doublets, the $Z_2$ symmetry forbids tree-level interactions of $\mathcal{N}_2$ other than the first term in Eq.~\eqref{eq:seesaw}. Following the spontaneous breaking of the $U(1)_L$ symmetry, the scalar field $\Phi$ can be expanded around its VEV as $\Phi = (v_\phi + \phi + i\chi)/\sqrt{2}$, and the RHN mass matrix takes the form
\bea
	M_R = \begin{pmatrix}
		0 & M_N \\
		M_N & 0
	\end{pmatrix},~~~~M_N = \frac{f v_{\phi}}{\sqrt{2}}
	\eea
which effectively leads to two exactly degenerate RHNs. Moreover, the CP even component of $\Phi$ also becomes massive at this stage while the imaginary component $\chi$ (Majoron) remains massless. 

To generate viable light neutrino masses, we additionally introduce the following dimension-five explicit $U(1)_L$-breaking operators, suppressed by a cutoff scale $\Lambda$:
\begin{eqnarray}
	-\mathcal{L}_{\sla{L}}&=& \frac{c_1}{2\Lambda}{\left[\Phi^2+(\Phi^*)^2\right]\overline{\mathcal{N}_1^C}\mathcal{N}_1}+\frac{c_2}{2\Lambda}{\left[\Phi^2+(\Phi^*)^2\right]\overline{\mathcal{N}_2^C}\mathcal{N}_2} \nonumber \\
	&+&\frac{\lambda_{\alpha 2}}{\Lambda}\overline{L_\alpha} \tilde{H}\mathcal{N}_2 (\Phi+\Phi^*)+h.c.,
\label{eq:dim5-explicit}	
\end{eqnarray} 
such that other neutrino Yukawa contribution ($y_{\alpha 2} = \lambda_{\alpha 2}\sqrt{2}v_{\phi}/\Lambda$) as well as RHN masses are induced after the $U(1)_L$ symmetry being spontaneously broken\footnote{We assume identical coefficients for $\Phi$ and $\Phi^\star$ in the higher-dimensional operators, enforcing a CP symmetry under $\Phi \to \Phi^\star$ \cite{Gross:2017dan}.}. As a result, following Eq. \eqref{eq:seesaw} and Eq. \eqref{eq:dim5-explicit}, the effective RHN mass matrix becomes 
\beeq
M_R^{\rm eff}=\begin{pmatrix}
	{\kappa} & M_N \\
	M_N & {\kappa}
\end{pmatrix},~~~~~\kappa=c\frac{v_\phi^2}{\Lambda},
\eeq
where $c_1 = c_2 =c$ is considered. 
This RHN mass matrix can now be trivially diagonalized to
 \begin{eqnarray}
      M_R^d = {\rm{diag}}[M_N+\kappa, M_N-\kappa], \label{RHN mass}
 \end{eqnarray}
by rotating through a unitary matrix $V$ as $V^* M_R V^\dagger = M_R^{d}$, where
\begin{eqnarray}
V = \frac{i}{\sqrt{2}}
\begin{pmatrix}
		1 & -1 \\
		i & i
	\end{pmatrix}.
\qquad
\label{V-matrix}
\end{eqnarray}
\noindent Consequently, this leads to the quasi-degenerate RHN states $N_{{1,2}}$, expressed as $\mathcal{N}_1=\frac{1}{\sqrt{2}}(N_{1}+i N_{2})~~{\rm and}~~\mathcal{N}_2=\frac{1}{\sqrt{2}}(N_{1}-iN_{2})$ with their respective masses and the mass splitting expressed as
\begin{align}
	M_{1,2}=\frac{fv_\phi}{\sqrt{2}} \pm \frac{v_\phi^2}{\Lambda}; ~~
	\Delta M^0=2\frac{v_\phi^2}{\Lambda},
	\label{mass-splitting}
\end{align}
where we put $c$ = 1 for simplicity. Additional higher-dimensional operators involving explicit $U(1)_L$ breaking term in the scalar sector as well as dimension-five operators that break the $Z_2$ symmetry, may also be present. Their implications for Majoron phenomenology and the domain-wall problem will be discussed later.


Once the electroweak symmetry is also broken, the light neutrino masses arise\footnote{The Weinberg operator is taken to be subdominant, ensuring that neutrino masses arise primarily from the seesaw mechanism.} via the type-I seesaw mechanism
\beeq
m_\nu\simeq m^D M_R^{-1}(m^{D})^T,
\eeq
with $m_{\alpha i}^D=y_{\alpha i} v/\sqrt{2}$ being the Dirac mass matrix (with $v=246$ GeV, the VEV of the SM Higgs boson and $i=1,2$), consisting of contributions from both the renormalisable and non-renormalizable terms. The Dirac mass matrix $m^D_{\alpha i}$, crucial for neutrino masses, mixing and leptogenesis, is fixed by neutrino oscillation data \cite{ParticleDataGroup:2024cfk} (we consider a normal mass hierarchy here). With two RHNs, the lightest neutrino remains massless and the Dirac mass matrix being of $3\times 2$ structure, its two columns $m_{\alpha i}^D$ can be expressed\footnote{ We note that ref.\cite{Batra:2023ssq} includes an additional factor of $e^{i\pi/2}$ in the Dirac columns of Eq.~\eqref{dirac_columns}. As this phase does not contribute to CP violation, it leaves the CP asymmetry unchanged and is consequently omitted here.}
in terms of  neutrino oscillation parameters and two additional Dirac mass parameters $m_{1,2}^D$ as \cite{Gavela:2009cd, King:2025eqv}
\begin{align}
m^D_{\alpha  1} &= m_1^D\left(\sqrt{1-\rho}\,U_{\alpha 2} + \sqrt{1+\rho}\,U_{\alpha 3}\right),\nonumber \\
m^D_{\alpha 2} &= m_2^D\left(\sqrt{1-\rho}\,U_{\alpha 2} - \sqrt{1+\rho}\,U_{\alpha 3}\right).
\label{dirac_columns}
\end{align}
\noindent Among the oscillation parameters, 
\begin{equation}
\rho= \frac{\sqrt{1+r}-\sqrt{r}}{\sqrt{1+r}+\sqrt{r}},
\end{equation}
with
\begin{equation}
r= \frac{|\Delta m^2_{\rm solar}|}{|\Delta m^2_{\rm atmos}|}=\frac{|\Delta m^2_{21}|}{|\Delta m^2_{32}|},
\end{equation}
\noindent and the matrix PMNS matrix $U$ is given by
 \begin{widetext}
\be
U=\left(
\begin{array}{ccc}
 c_{12} c_{13} & c_{13} s_{12} & e^{-i \delta_{CP} } s_{13} \\
 -c_{23} s_{12}-e^{i \delta_{CP} } c_{12} s_{13} s_{23} & c_{12} c_{23}-e^{i \delta_{CP} } s_{12} s_{13} s_{23} & c_{13} s_{23} \\
 s_{12} s_{23}-e^{i \delta_{CP} } c_{12} c_{23} s_{13} & -e^{i \delta_{CP} } c_{23} s_{12} s_{13}-c_{12} s_{23} & c_{13} c_{23} \\
\end{array}
\right)\times {\rm{diag}}(1,e^{i\alpha/2},1),
\label{pmns}
\ee
 \end{widetext}
having $\theta_{12},\theta_{23},\theta_{13}$ as the mixing angles ($c_{ij}=\cos\theta_{ij},s_{ij}=\sin\theta_{ij}$).
Here the Dirac CP phase is $\delta_{CP}$ and the Majorana CP phase is denoted by $\alpha$. With this parametrisation, the light neutrino masses satisfy
\begin{equation}
m_1 = 0,~
m_2 = \frac{m_1^D m_2^D}{M_N}(1-\rho),~
m_3 = \frac{m_1^D m_2^D}{M_N}(1+\rho).
\end{equation}
Therefore, the light neutrino mass matrix in this parametrisation\footnote{The chosen parametrisation is particularly useful for two quasi-degenerate RHNs. It coincides with the Casas–Ibarra parametrisation \cite{Casas:2001sr} for CI rotation angles $0$ or $\pi$, rendering the rotation matrix fixed and real; hence, CP asymmetry arises solely from the Dirac and Majorana phases, as in the parametrisation adopted here.} depends on the interplay between $m_i^D$ satisfying 
\beeq
\frac{m^D_{1}m^D_{2}}{M_N}\approx 0.03 \ {\rm eV}.
\label{product}
\eeq
With the neutrino mass structure and the closely degenerate RHN spectrum specified above, forming a quasi-Dirac fermion, we now investigate the possibility of generating the observed baryon asymmetry through Higgs-decay leptogenesis. In this scenario, finite-temperature effects enable the decay ($H\to L N_i$) in the early Universe, while the quasi-degeneracy of the RHNs allows a resonant enhancement of the CP asymmetry. In the following section, we discuss the relevant thermal effects and the resulting lepton asymmetry, focusing on the interplay between the symmetry-induced vacuum mass splitting and the temperature-dependent contributions to the RHN dynamics.

  


\section{Low-scale Leptogenesis from Higgs decay}
\label{sec.leptogenesis}

\begin{figure}[tb!]
    \centering
    \includegraphics[width=0.45\linewidth]{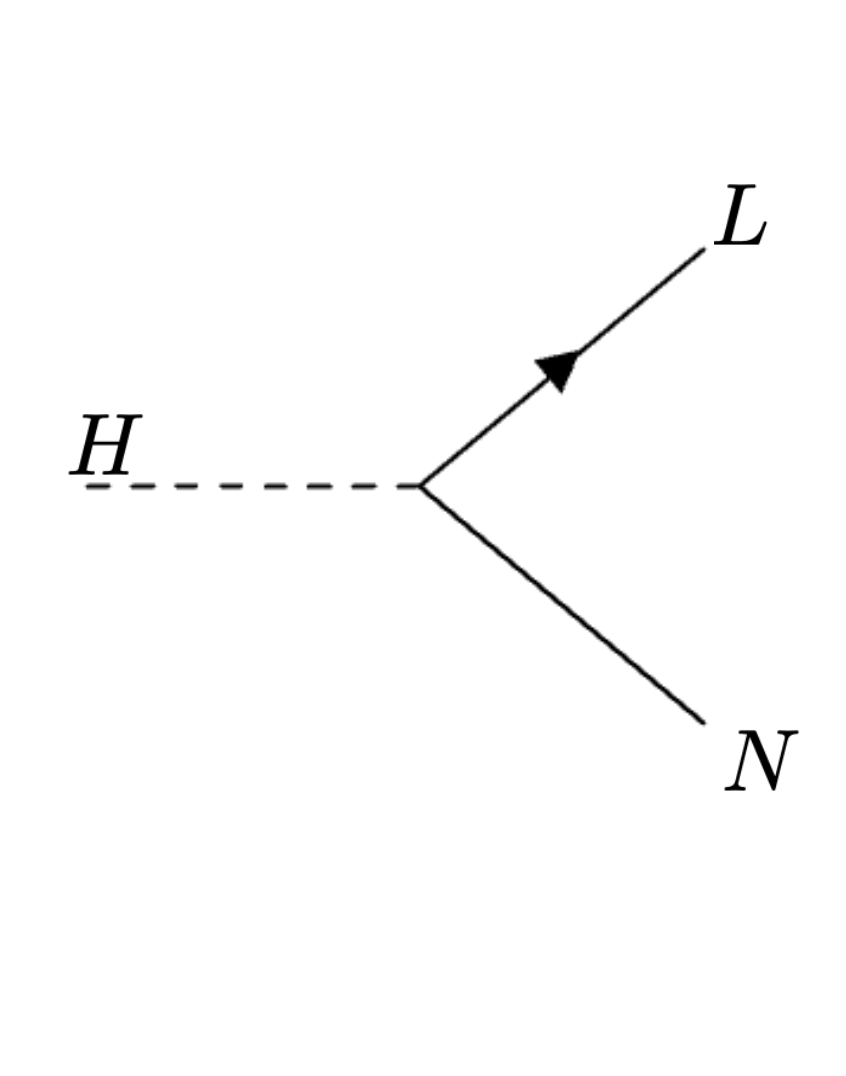}
    \includegraphics[width=0.6\linewidth]{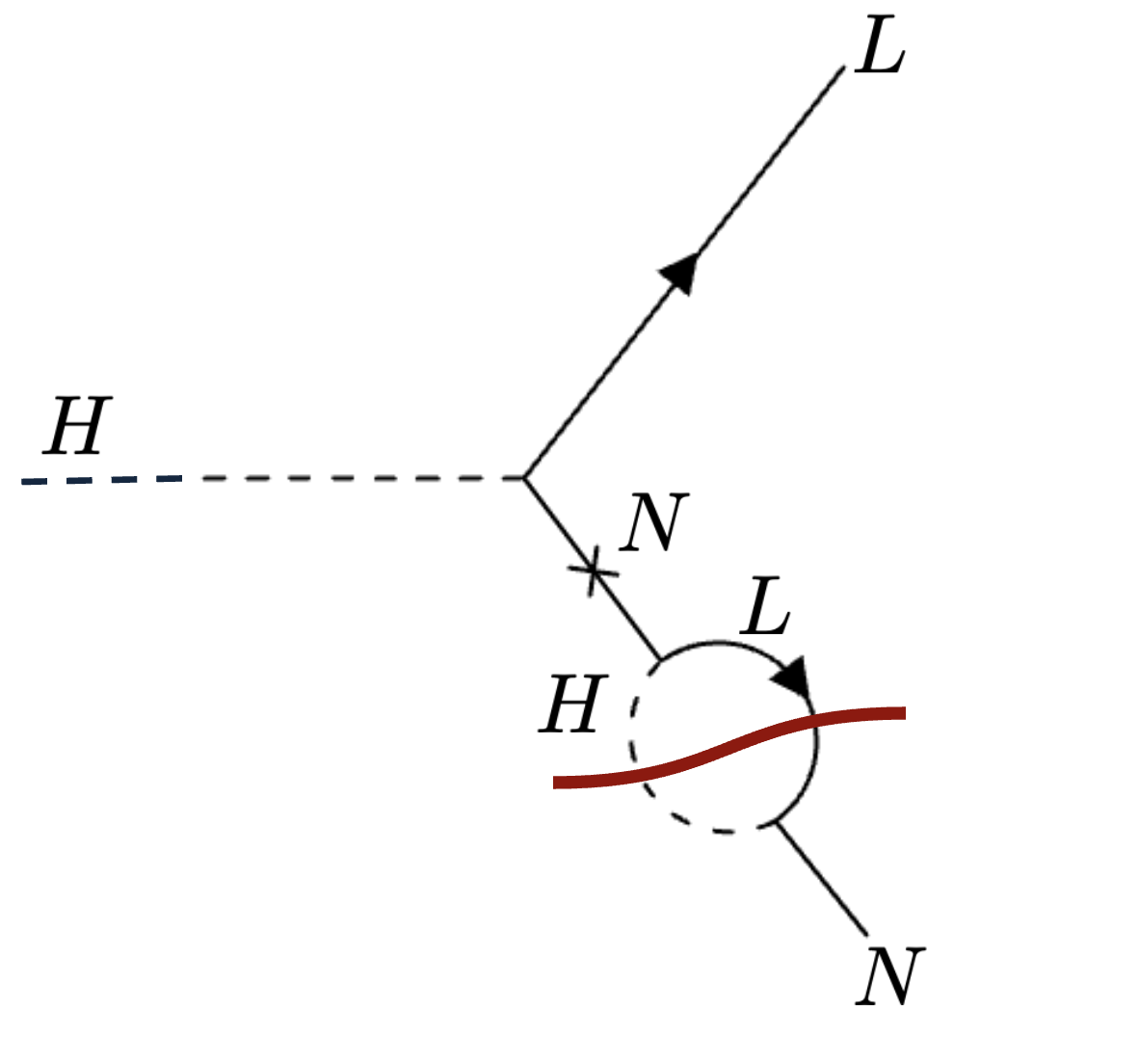}
    \caption{Tree-level and self-energy diagrams contributing to the generation of the asymmetry through Higgs decay, $H\to LN$. The thermal cut is highlighted in dark red.}
    \label{fey_dig}
\end{figure}

Here, we consider a mechanism of lepton asymmetry generation through SM Higgs boson decay, as proposed in Ref. \cite{Hambye:2016sby}, which differs from conventional thermal leptogenesis based on RHN decay. In the conventional scenario, the lepton asymmetry is generated through the decay of RHNs: $N\to L H$. However, for the RHN masses of interest to us, {\it i.e.}, $\mathcal{O}$(GeV), this mechanism is  inefficient. In particular, the CP asymmetry in hierarchical $N\to L H$ decays is bounded from above by the Davidson-Ibarra limit \cite{Davidson:2002qv}, which scales linearly with $M_N$ and requires $M_N\gtrsim 10^9$ GeV for standard thermal leptogenesis, while the decay becomes kinematically forbidden at temperatures relevant for the broken electroweak phase. We therefore consider the alternative possibility of generating the lepton asymmetry from the decay of the thermal Higgs boson, $H\to L N$, which becomes relevant around the electroweak scale, particularly in the temperature window $T_c~(\sim 160$ GeV) $\gtrsim T \gtrsim T_{\rm sph}~(\sim 131.7$ GeV), where $T_c$ and $T_{\rm sph}$ denote the EWSB critical temperature and sphaleron freeze-out temperature, respectively. Unlike conventional $N$-decay leptogenesis, the GeV-scale RHNs are produced through Higgs decays, and the CP asymmetry arises from the interference between the tree-level and loop-level contributions, as shown in Fig. \ref{fey_dig} , to the Higgs decay amplitude.

It is important to note that although the decay channel $H\to L N$ is kinematically allowed below $T_c$, the one-loop diagram does not develop an absorptive (imaginary) part in vacuum (in absence of thermal plasma). For an absorptive part to arise, the intermediate particles in the loop should simultaneously be on-shell. In the present case, the relevant intermediate state is the $H,L$ pair, for which the on-shell condition: $M_N>m_H+m_L$ can't be satisfied. Here, $m_H$ and $m_L$ denote the Higgs and lepton-doublet masses, respectively \cite{Giudice:2003jh}. Consequently, the loop remains purely real at $T=0$, and no CP asymmetry is generated in vacuum.
The situation however changes in presence of thermal plasma, {\it{i.e.}} taking into account the thermal effect, as the Higgs and lepton fields are in thermal plasma. The relevant loop contribution can then acquire an absorptive part through thermal processes involving particles from the bath \cite{Giudice:2003jh}. For example, the corresponding thermal cut can be interpreted as the scattering process $N+H_{\rm bath}\to L$, in which the bath Higgs participates as an on-shell state. Such thermal processes provide the required absorptive contribution to the loop, which is absent at $T=0$, and hence allow a non-zero CP asymmetry in Higgs decay \cite{Hambye:2016sby}.

The CP asymmetry in this framework is therefore intrinsically thermal in origin as discussed above. Considering further that the two RHNs are closely degenerate, the CP asymmetry can be expressed in terms of $x=M_N/T$ as the following \cite{Hambye:2016sby}:
\begin{equation}
	\epsilon_{\rm CP}(x) \ = \ I_1 \;    \frac{2~\Delta M^0 ~\Gamma_T(x)}{4\,\Delta M(x)^2 + \Gamma_T(x)^2} \;,
	\label{eq:asym}
\end{equation}
where,
\beeq
I_1 = \frac{{\rm Im}\left[(y_\nu y_\nu^\dagger)^2_{12}\right]}{\left(|y_\nu y_\nu^\dagger|_{11}~|y_\nu y_\nu^\dagger|_{22}\right)}.
\eeq
The Dirac coupling matrix $y_{\nu}$ in the diagonal basis of the RHN is obtained from 
\begin{equation}
\tilde{m}_D \equiv y_{\nu} v/\sqrt{2} = m_D \,V^\dagger, 
\label{y-neu}
\end{equation}
using Eq. \eqref{V-matrix} and $\Delta M(x)$ is defined as 
\begin{equation}
    \Delta M(x)=\Delta M^0+\Delta M^T(x),
\end{equation}
with 
$\Delta M^T(x)$ implying the thermal corrections to the RHN bare mass-splitting $\Delta M^0$, given by
\begin{align}
	\Delta M^T(x) \ &\simeq \ \frac{\pi}{4 x^2} \, \Gamma_{22} \; \sqrt{\left(1 - \frac{\Gamma_{11}}{\Gamma_{22}}\right)^2 + 4 \,\frac{|\Gamma_{12}|^2}{\Gamma_{22}^2}} \nonumber\\
	& \equiv \ \frac{\pi}{4 x^2} \, \Gamma_{22} \, f \;.
	\label{eq:th-masses}
\end{align}
Here, $\Gamma_{ij}\simeq M_N (|y_\nu|^2_{ij}/8 \pi)$ denotes the decay width of the RHNs. The 
$\Gamma_T(x)$ in Eq. \eqref{eq:asym} refers to the temperature dependent decay width, where the temperature dependence can be extracted as $\Gamma_T(x)=\Gamma_{22}\gamma(x)$ with $\gamma(x)$ expressed as \cite{Frossard:2012pc}
	$\gamma(x) \equiv \frac{p \cdot L_\rho(q)}{p \cdot q}$.
Here $p$ and $q$ denote the four-momenta of the charged lepton and RHN, respectively, and $L_\rho(q)$ is the absorptive part of the RHN self-energy. It encodes the thermal phase-space and statistical effects through distribution functions of the plasma particles. The detailed computation of these thermal contributions can be found in Ref. \cite{Frossard:2012pc}, which can be utilized to rearrange the form of $\epsilon_{\rm CP}(x)$ in Eq. \eqref{eq:asym} as \cite{Hambye:2016sby}
\beeq
\epsilon_{\rm CP}(x)= I_1\frac{2\frac{\Delta M^0}{\Gamma_{22}}\gamma(x)}{\left[2\frac{\Delta M^0}{\Gamma_{22}}+\frac{\pi}{4x^2}f\right]^2+\gamma(x)^2}.
\eeq
In the high temperature regime $(x\ll 1)$, relevant for GeV-scale RHNs, the thermal factor can be approximated as $\gamma(x)\simeq 50$, as estimated in Ref. \cite{Hambye:2016sby}. Since we plan to evaluate the lepton asymmetry up to the sphaleron freeze-out temperature, this condition is well suited for our interest of RHN mass, and hence, we continue with this approximation.

One can readily see from Eq. \eqref{eq:asym} that the CP asymmetry is maximized in the resonant regime $\Delta M_N(x)\sim \Gamma_T (x)$, which plays a key role in achieving successful leptogenesis at low scales. Such resonant enhancement of the CP asymmetry is clearly regulated by the thermal effects here, which is included both in net RHN mass splitting $\Delta M_N(x)$ as well as in the decay width of the RHNs 
$\Gamma_T(x)$. With these, we then solve the Boltzmann equations to study the evolution of the RHN and the lepton asymmetry, which can be expressed in terms of the comoving number densities $Y_i=n_i/\mathcal{S}$ as:
\bea
\frac{dY_{N_i}}{dx}&=& -\frac{1}{\mathcal{H}x\mathcal{S}} \left[\frac{Y_{N_i}}{Y_{N_i}^{\rm eq}}-1\right]\gamma_D,\nonumber\\
\frac{dY_{B-L}}{dx}&=& -\sum_{i}\frac{1}{\mathcal{H}x\mathcal{S}} \left[\epsilon_{\rm CP}\left(\frac{Y_{N_i}}{Y_{N_i}^{\rm eq}}-1 \right)+\frac{Y_{L}}{2Y_{l}^{\rm eq}}\right]\gamma_D.\nonumber\\
\label{BE_asy}
\eea
Here, $\mathcal{S}=\frac{2\pi^2}{45}g_\star^\mathcal{S}(T)T^3$ and $\mathcal{H}=1.66\sqrt{g_\star^\rho(T)}\frac{T^2}{M_{\rm Pl}}$ are the entropy density and the Hubble expansion rate of the radiation dominated Universe, respectively. The $Y^{{\rm eq}}_X$ in Eq. \eqref{BE_asy} follows
\beeq
Y_{X}^{{\rm eq}} (T)= \frac{45 g_X}{4\pi^4 g^\mathcal{S}_{\star}(T)}\left( \frac{m_X}{T}\right)^2 K_2\left(\frac{m_X}{T} \right),
\eeq
which refers to the equilibrium yield of a given particle species $`X$' with $m_X$ and $g_X$ being its mass and internal d.o.fs, respectively.
$\gamma_D$ in Eq. \eqref{BE_asy} signifies the total decay rate, given by $\gamma_D \ = \ \gamma_{N\to L H} \,\theta(m_N - m_H - m_L) \; + \; \gamma_{H\to NL} \,\theta(m_H - m_N - m_L)$ with $\gamma_{N\to L H}$ and $\gamma_{H \to NL}$ implying the thermally averaged decay rates, defined as
\begin{align}
	\gamma_{N\to L H}  &= \frac{M_N^3}{\pi^2 x} \, K_1(x) \, \Gamma_{N\to L H}   \;,\\
	\gamma_{H\to N L}  &= \frac{m_H^2 \, M_N}{\pi^2 x} \, K_1\bigg(\frac{m_H}{M_N} x\bigg) \, 2\, \Gamma_{H\to N L} \;,
\end{align}
where
\begin{align}
	\Gamma_{N\to L H}  &=  \frac{M_N}{8\pi} \, y_\nu y_\nu^\dag \;   \lambda^{\frac{1}{2}}\!(1, a_H, a_L) \, (1- a_H + a_L) \;,\\
	\Gamma_{H \to NL}  &=  \frac{M_N}{8 \pi} \, y_\nu y_\nu^\dag \; \lambda^{\frac{1}{2}}\!(1, a_H, a_L) \,  \frac{a_H - a_L - 1}{2 \, a_H^{3/2}} \;.
\end{align}
Here, $a_X \equiv (m_X(T)/M_N)^2$ and $\lambda(x,y,z)=x^2+y^2+z^2-2xy-2xz-2yz$ \cite{Frossard:2012pc}. We use these equations to present the evolution of the lepton asymmetry, followed by a detailed analysis and parameter space scan in the later sections.

\section{Comparison with ARS mechanism}
\label{ARS}


In the same RHN mass regime we are interested in, $M_N\sim\mathcal{O}(\mathrm{GeV})$, another viable mechanism for low-scale leptogenesis is the ARS mechanism \cite{Akhmedov:1998qx}. It is therefore pertinent to compare this mechanism with the Higgs-decay leptogenesis scenario considered in this work before proceeding for phenomenological discussion as the two rely on fundamentally different origins of the CP asymmetry and mechanisms for generating the baryon asymmetry.

In the ARS mechanism, the baryon asymmetry originates from the CP-violating coherent oscillation of nearly degenerate RHNs. These RHNs are initially assumed to be absent in the bath and are gradually populated out-of-equilibrium via feeble Yukawa couplings. Because the total lepton number $L=L_{\rm SM}+L_{\rm N}$ remains conserved during the production and oscillation, these oscillations generate opposite lepton asymmetries in different SM lepton flavors, resulting in a net lepton asymmetry to be zero. A non-zero active lepton asymmetry then emerges through flavor-dependent washout processes that erase individual flavor asymmetries at different rates, leaving a non-zero total asymmetry in the SM sector. Electroweak sphaleron subsequently convert this surviving lepton asymmetry into a baryon asymmetry. Thus, ARS mechanism relies crucially on coherent RHN oscillations and flavor-dependent washout, whereas the mechanism considered here does not require such RHN oscillations to induce CP asymmetry.

Another crucial distinction between the two scenarios lies in their characteristic temperature scales and sensitivity to the RHN mass splitting. In the ARS framework, the characteristic temperature at which the quantum oscillations between two RHNs commence becomes important, which is governed by the condition involving the RHN oscillation frequency $\omega_{\rm osc}\simeq \frac{\Delta M^2}{2T}, \Delta M^2 \equiv M_2^2-M_1^2$,  
with Hubble parameter $\mathcal{H}(T)$ via $\omega_{\rm osc}\sim \mathcal{H}(T)$, resulting \cite{Drewes:2017zyw,Escudero:2021rfi}
\beeq
T_{\rm osc}\sim 10^5 ~{\rm GeV} \left(\frac{M_N}{10~{\rm GeV}}\right)^{2/3}\left(\frac{\Delta M^0}{10^{-4}M_N}\right)^{1/3},
\eeq
for quasi degenerate RHNs.
It immediately implies that for a given $M_N$, the onset of oscillations is sensitive to the RHN mass degeneracy. As a result the requirement of the oscillations to occur before sphaleron freeze-out (i.e. $T_{\rm osc}>T_{\rm sph}$) imposes a lower bound on $\Delta M^0$. 

\begin{figure}[h]
	\includegraphics[width=1\linewidth]{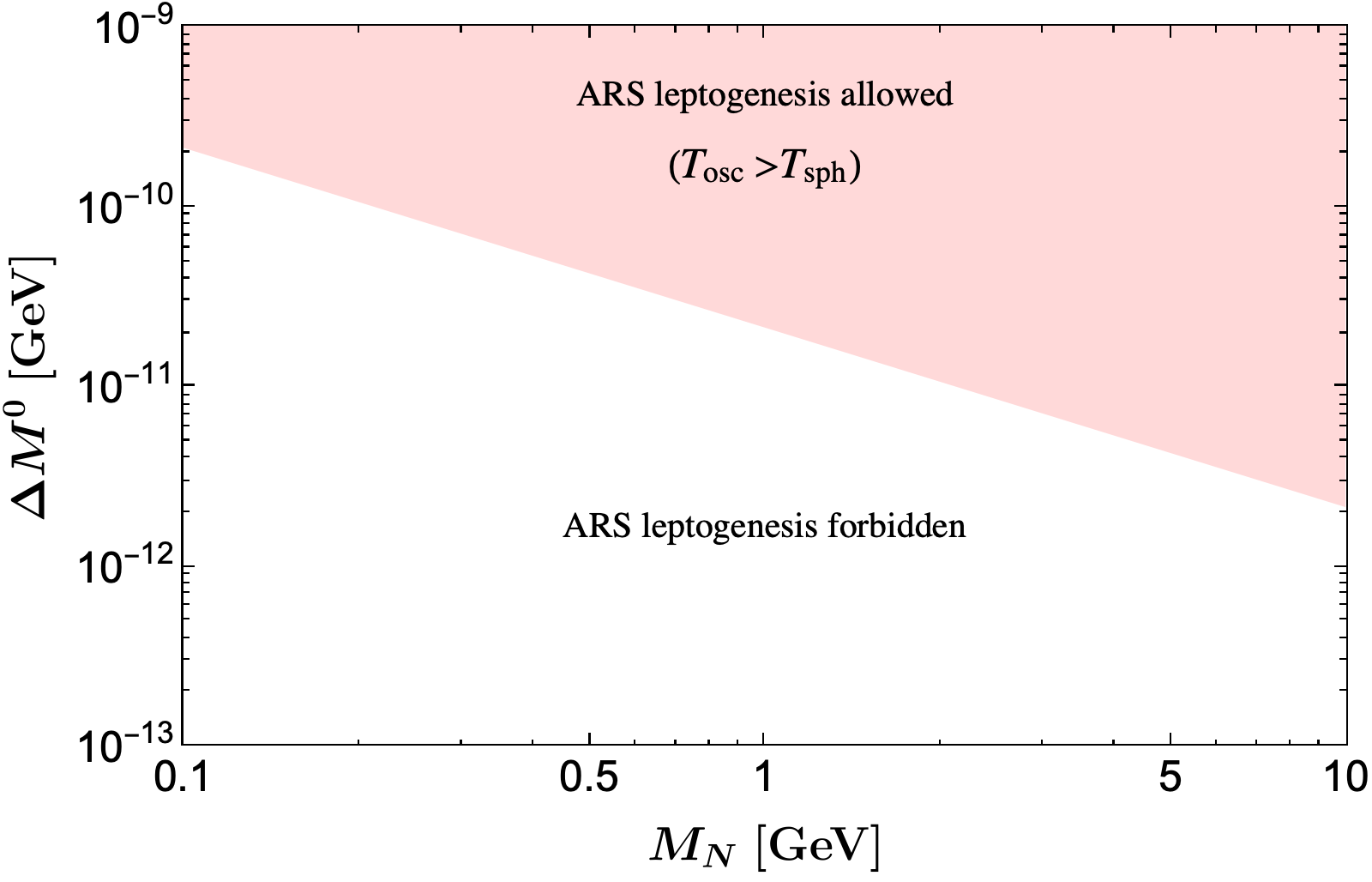}
	\caption{Region of parameter space in the $\Delta M^0$--$M_N$ plane where only Higgs-decay leptogenesis is viable (white region). The pink shaded region indicates where ARS leptogenesis can also contribute, with its boundary corresponding to $T_{\rm osc}=T_{\rm sph}$.  }
	\label{fig:ARScomp}
\end{figure}

The existence of such lower bound on $\Delta M^0$ suggest that, below this splitting, the RHNs oscillate too late (after electroweak sphaleron have already decoupled), rendering the ARS mechanism completely incapable of producing a net baryon asymmetry. This provides an important distinction for the parameter space explored in the present work. As we will notice in the following sections, the vacuum mass splitting in our framework is generated naturally through Eq. \eqref{mass-splitting}, and can therefore be much smaller than the splitting typically relevant for ARS oscillations. As indicated in Fig. \ref{fig:ARScomp} (also in Figs. \ref{fig:DM0-YB1} and \ref{fig:DM0-YB2}), only in the pink-shaded region, where $T_{\rm osc}>T_{\rm sph}$, ARS mechanism can be operative, although determining whether it competes with or dominates over the Higgs decay mechanism would require a dedicated analysis involving the quantum kinetic equations which is beyond the scope of this work.
Therefore, the novelty of this work lies in realizing such low RHN mass-splitting regime, establishing Higgs decay as the sole and dominant mechanism for leptogenesis in this parameter space. Moreover, the same symmetry-breaking structure responsible for the mass splitting is connected to the formation and subsequent decay of DWs, providing a link between the low-scale leptogenesis mechanism and the stochastic gravitational-wave signal discussed in the Section \ref{sec:GW}.

\section{Results and Analyses}
\label{results}

\begin{figure}[tb!]
    \centering
    \includegraphics[width=1\linewidth]{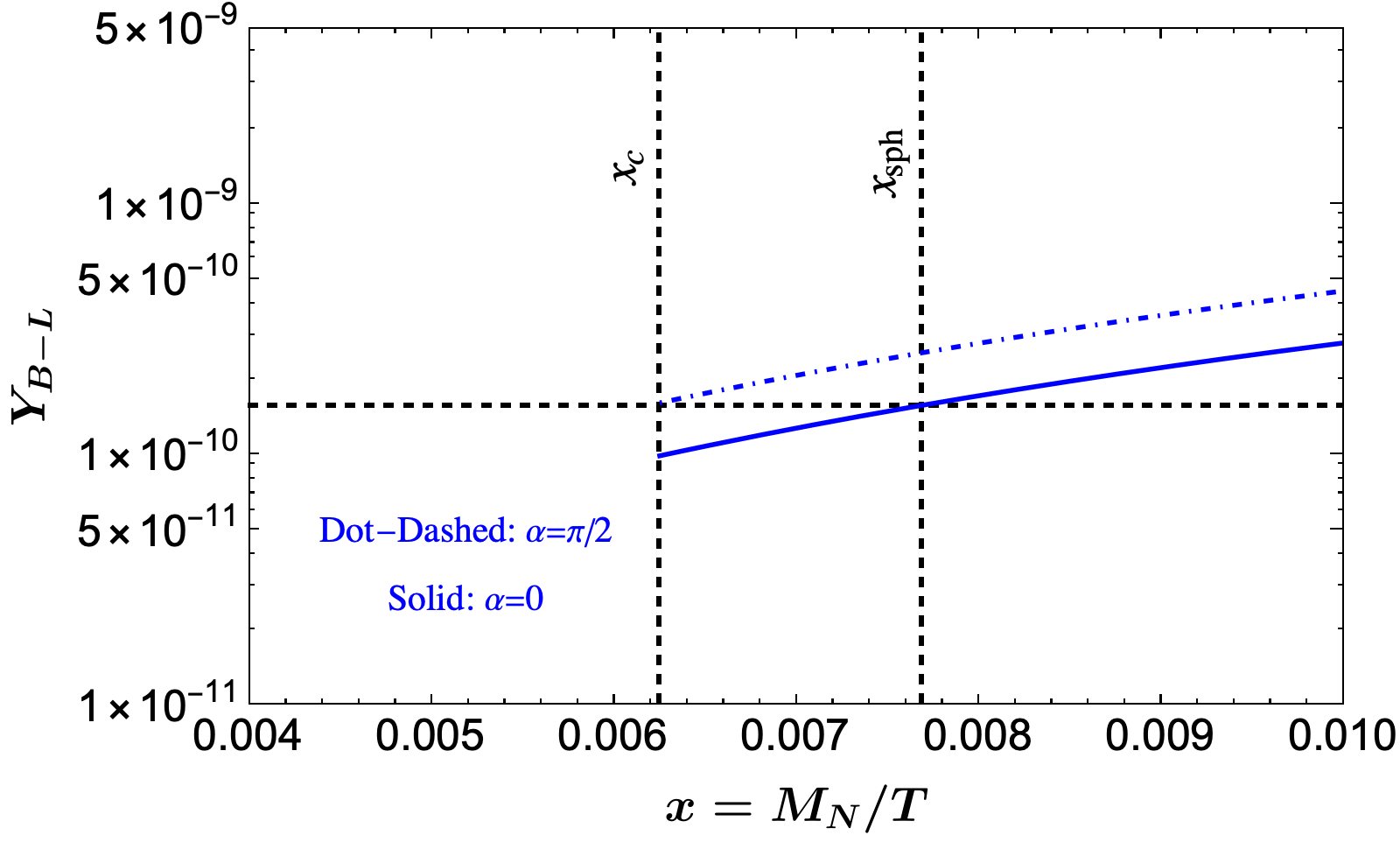}
    \caption{Evolution of the $B-L$ asymmetry (dot-dashed and solid blue lines) for the benchmark point: $M_N=1$ GeV, $\Delta M^0=9\times10^{-10}$ GeV, $m_1^{D}=10^{-5}$ GeV. The vertical
    dashed lines denote the EWSB critical and sphaleron freeze-out temperatures,
    respectively, while the horizontal dashed line indicates the amount of $B-L$
    asymmetry required to reproduce the observed baryon asymmetry of the
    Universe.}
    \label{fig:BP1}
\end{figure}

We begin by illustrating the evolution of the $B-L$ asymmetry in
Fig.~\ref{fig:BP1} for the benchmark parameter set: $M_N = 1~{\rm GeV}$,
$\Delta M^0 = 9\times 10^{-10}~{\rm GeV}$, $m_1^D = 10^{-5}$~GeV, and two
different values of the Majorana phase, $\alpha =  0,\,
\frac{\pi}{2}$. The plot is obtained by solving the coupled Boltzmann equations
in Eq.~\eqref{BE_asy} with the initial conditions $Y_N(x_c) = 0$ and
$Y_{B-L}(x_c) = 0$, where $x_c = M_N/T_c$. This choice reflects the fact
that, in the present scenario, dominant asymmetry production begins only at the critical
temperature of electroweak symmetry breaking, $T_c$.

\noindent As is evident from Fig.~\ref{fig:BP1}, the asymmetry is generated
predominantly in the temperature range $T_c \gtrsim T \gtrsim T_{\rm sph}$,
where the Higgs decay channel remains active. We include thermal mass
corrections to the SM fields of the form $m_i^2(T) \sim m_i^2 + c_i T^2$
\cite{Giudice:2003jh}, which also allow the relevant process to proceed
slightly above $T_c$, albeit less efficiently due to phase-space suppression.
Although the $B-L$ asymmetry continues to be produced down to much lower
temperatures, it can be converted into a baryon asymmetry only while the
sphaleron processes remain in equilibrium, once they freeze out at
$T_{\rm sph}$, only the asymmetry present at this temperature contributes to
the final baryon abundance. For the benchmark parameters above, the observed
BAU is thus successfully reproduced at $T_{\rm sph}$ (vertical black dashed
line) through the relation $Y_B = (28/79)\,Y_{B-L}(x_{\rm sph}) \simeq
8.75\times 10^{-11}$.

We further note that the lepton asymmetry yield is insensitive to the choice of
the initial temperature, provided it lies above the electroweak scale, since
the asymmetry is generated dominantly at later times. We assume that the RHNs
are initially absent and remain out of equilibrium owing to their small Yukawa
couplings of $\mathcal{O}(10^{-8})$, as estimated using Eq. \eqref{y-neu} 
while the best fit values of neutrino oscillation parameters are considered. In this respect, and unlike conventional
leptogenesis, the Sakharov out-of-equilibrium condition is satisfied here by
the RHN produced in the final state, whose departure from equilibrium is
ensured by these suppressed Yukawa interactions.

\begin{figure}[tb!]
	\centering
	\includegraphics[width=1\linewidth]{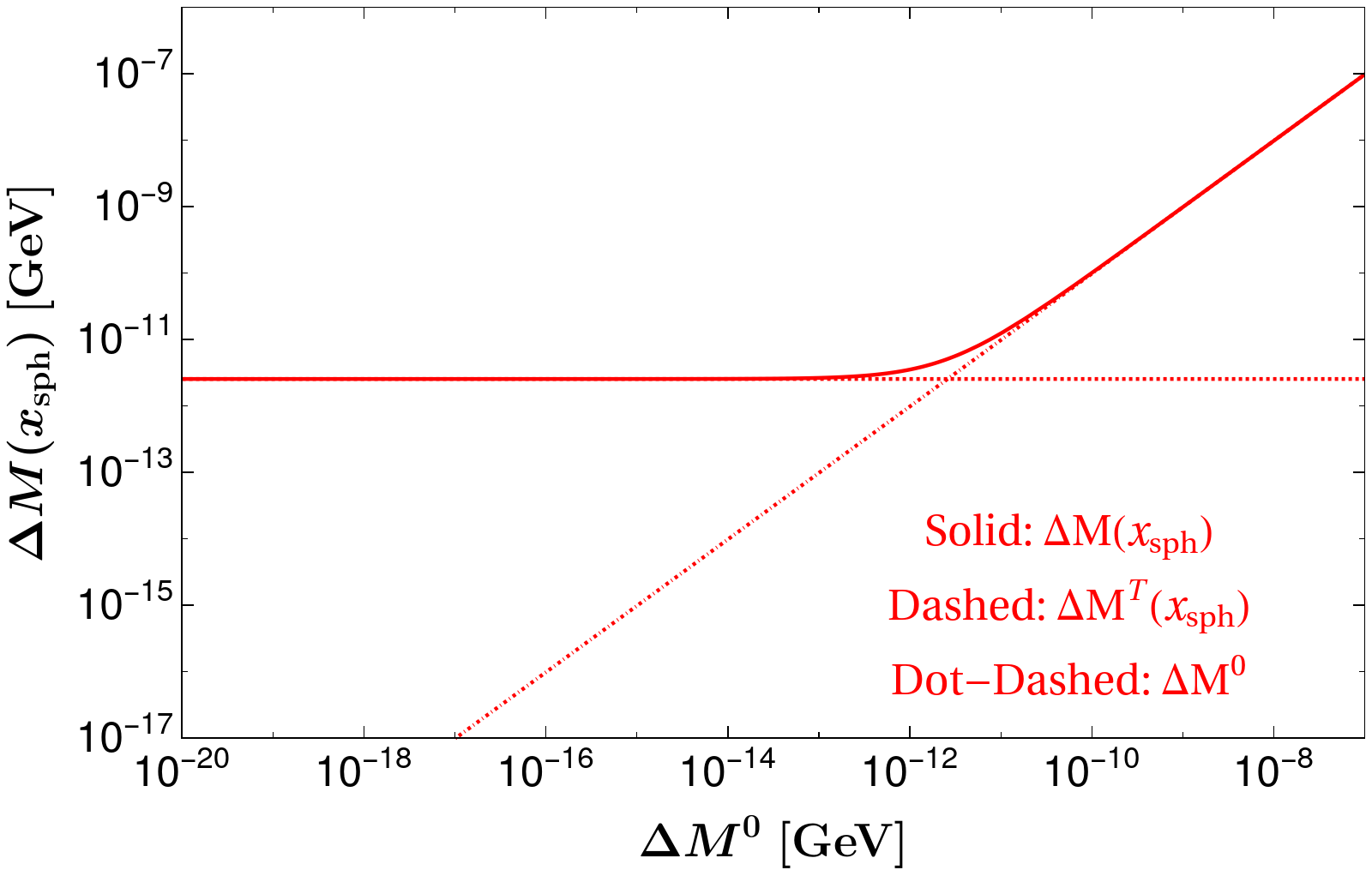}
	\caption{Comparison between temperature dependent and temperature-independent RHN mass splitting as a function of $\Delta M^0$ for $M_N=1$ GeV, $m^D_1=10^{-5}$ GeV and $\alpha=0$. Red dashed line indicates the thermal correction to the RHN mass splitting, evaluated at $x_{\rm sph}$ ($\Delta M^T (x_{\rm sph})$), red dot-dashed line corresponds to $\Delta M^0$ and the red solid line stands for the total RHN mass splitting at $x_{\rm sph}$ ($\Delta M (T_{\rm sph})$).}
	\label{fig:dM-all}
\end{figure}

\begin{figure}[tb!]
	\centering
	\includegraphics[width=1\linewidth]{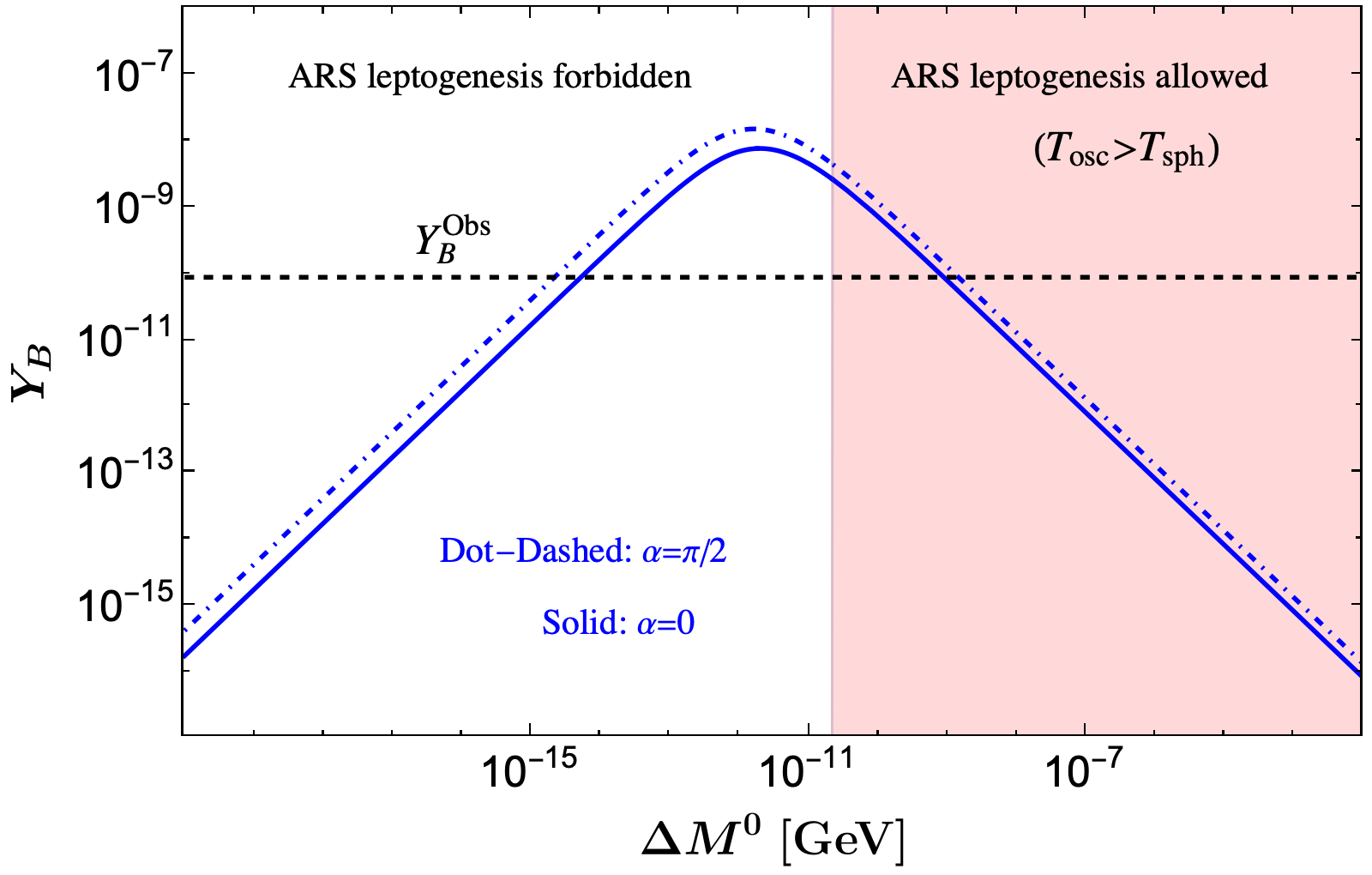}
	\caption{Baryon asymmetry $(Y_{B})$ at $T_{\rm sph}$ as a function of $\Delta M^0$ for benchmark values of $M_N=1$ GeV and $m^D_1=10^{-5}$ GeV for two different values of Majorana phases. The horizontal black dashed indicates the correct $Y_B$ to obtain the observed BAU.}
	\label{fig:DM0-YB1}
\end{figure}

Another key aspect of this low-scale leptogenesis is the zero-temperature RHN mass splitting $\Delta M^0$, emerging directly from the Lagrangian in Eq. \eqref{eq:dim5-explicit}. It is evident from the numerator of Eq. \eqref{eq:asym} that in the limit $\Delta M^0\to 0$, the CP asymmetry vanishes. Although we treat it as a free parameter so far, interestingly, such mass splitting arises naturally $\Delta M^0 = 2 v_\phi^2/\Lambda$ as discussed in the context of Eq. \eqref{mass-splitting}. 
Next we explore the dependence of baryon asymmetry on $\Delta M^0$. As shown in Fig. \ref{fig:DM0-YB1}, the correct BAU can also be achieved for different values of $\Delta M^0$ for a fixed $M_N$, $m^D_1$ and $\alpha$. This behaviour arises from the interplay between the vacuum and thermal contributions to the RHN mass splitting entering $\epsilon_{\rm CP} (x)$, as illustrated in Fig. \ref{fig:dM-all}, where the total mass splitting $\Delta M$ is evaluated at $T=T_{\rm sph}$, since most of the relevant $B-L$ asymmetry is produced at this temperature which contributes into BAU. 

\begin{figure}[tb!]
	\centering
	\includegraphics[width=1\linewidth]{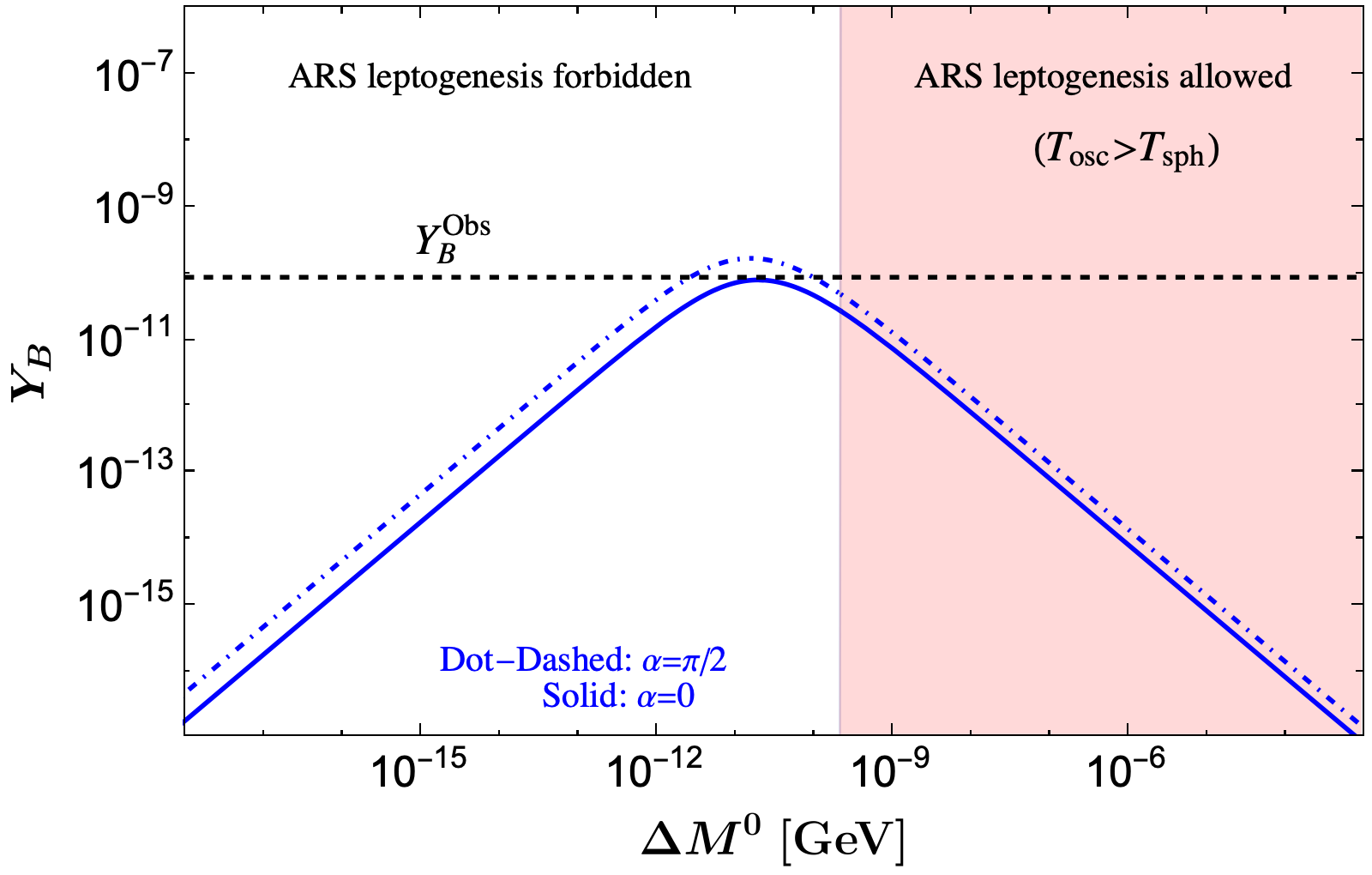}
	\caption{Baryon asymmetry $(Y_{B})$ at $T_{\rm sph}$ as a function of $\Delta M^0$ for benchmark values of $M_N=0.1$ GeV and $m^D_1=10^{-5}$ GeV for two different values of Majorana phases. The horizontal black dashed indicates the correct $Y_B$ to obtain the observed BAU.}
	\label{fig:DM0-YB2}
\end{figure}
 It can be seen from  Fig. \ref{fig:dM-all} that $\Delta M^T$ (evaluated at $T_{\rm sph}$) remains approximately constant at $\sim 2.57\times 10^{-12}$ GeV (indicated by the red dashed line). For $\Delta M^0\lesssim 2.57\times 10^{-12}$ GeV, the total splitting (red line) is dominated by the thermal contribution, $\Delta M(x)\approx \Delta M^T(x)$ and hence follows the red dashed curve. In this region, one can see from Eq. \eqref{eq:asym} that $\epsilon_{\rm CP}$ (and consequently $Y_{B-L}$) scales proportional to $\Delta M^0$, leading to an increase in $\epsilon_{\rm CP}$ ($Y_{B-L}$) with increasing $\Delta M^0~( \lesssim  2.57\times 10^{-12}~{\rm GeV})$ (see Fig. \ref{fig:DM0-YB1}). On the other hand, for $\Delta M^0\gtrsim  2.57\times 10^{-12}$ GeV, the vacuum contribution dominates, $\Delta M(x)\approx \Delta M^0$, and the total splitting follows the red dot-dashed curve. In this case, the CP asymmetry effectively becomes inversely proportional to $\Delta M^0$, resulting in a decrease in $\epsilon_{\rm CP}$ (and $Y_{B-L}$) with increasing $\Delta M^0$. In Fig. \ref{fig:DM0-YB2}, we do a similar analysis but for a $M_N=0.1$ GeV while keeping the same value of $m^D_1$. As expected, reducing $M_N$ to a lower value reduces the produced asymmetry. For $\alpha=0$, there exist only one value of $\Delta M^0$ for which the correct baryon asymmetry can be obtained.


\section{Experimental Prospects}
\label{sec:scan}
    \begin{figure*}[]
	\includegraphics[width=0.85\linewidth]{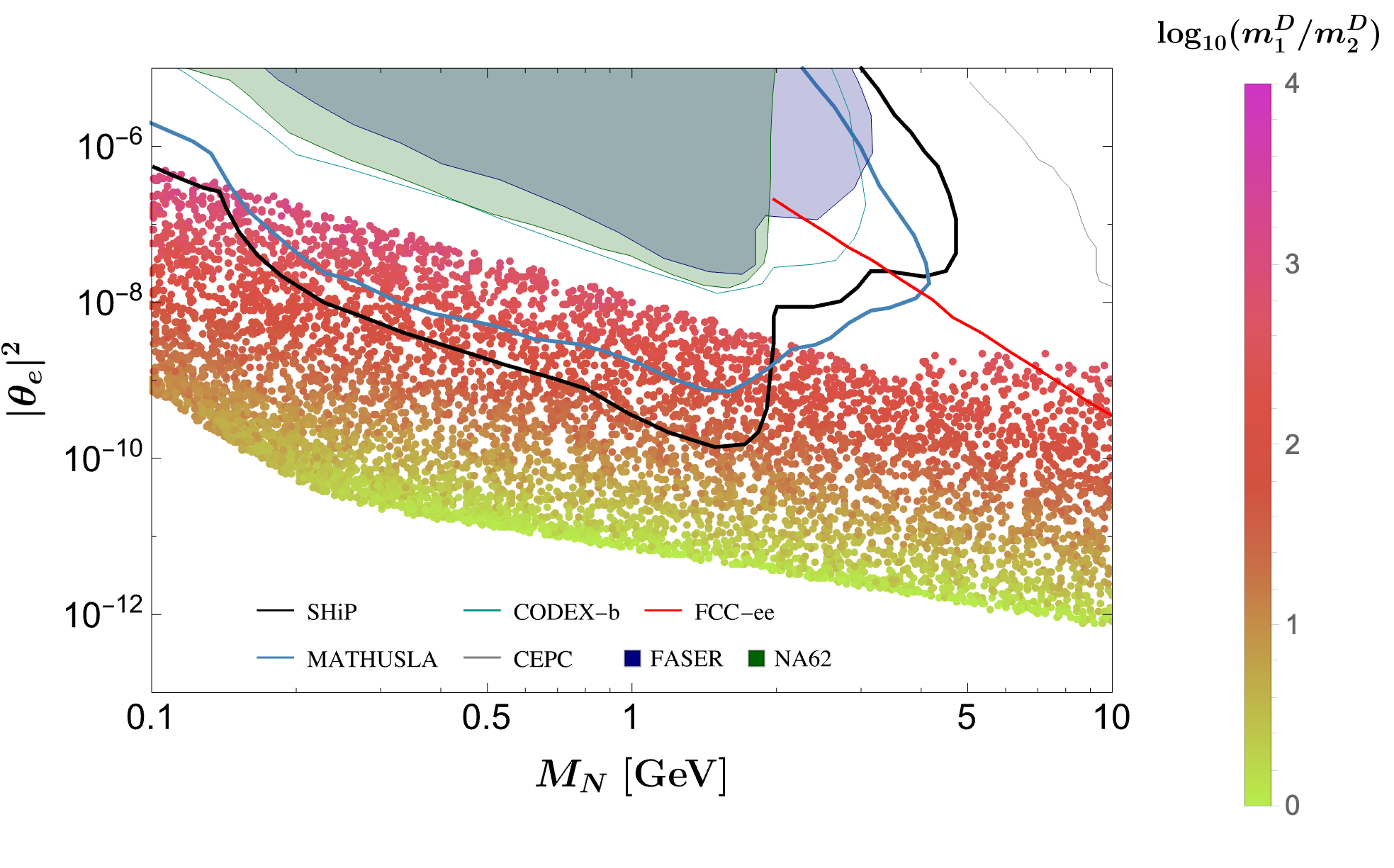}
	\caption{Present (shaded region) and projected (coloured contours) bounds on the active-sterile mixing parameter $|\theta_{e}|^2$ as a function of $M_N$ and parameter space in our framework. For generating the data points we consider $0<\alpha<2\pi$ and choose $\Delta M^0$ in such a way that no contributions come from ARS leptogenesis ($T_{\rm osc}<T_{\rm sph}$ is always satisfied). The heat map of points shows the ratio of Dirac mass parameters.}
	\label{fig:ship}
\end{figure*}

We now investigate the experimental prospects of our scenario by analyzing the active-sterile neutrino mixing $(\theta_\alpha)$ as a function of the RHN mass scale $M_N$, on which experimental bounds are mostly set. The RHN mass scale of our interest, $\mathcal{O}(0.1)-\mathcal{O}(10)$ GeV (below TeV scale), existing bounds arise from collider and beam dump experiments, while future facilities such as SHiP and FCC-ee are expected to significantly improve the sensitivity in the GeV mass range. In the present framework, the active-sterile neutrino mixing can be expressed as 
\beeq
|\theta_{\alpha}|^2 
\equiv  |\theta_{\alpha 1}|^2+ |\theta_{\alpha 2}|^2\approx 
\frac{({{m}}^D_{\alpha 2})^2}{M_N^2}+  \frac{({{m}}^D_{\alpha 1})^2}{M_N^2}.
\label{mixing}
\eeq	
This relation (in the limit of degenerate RHNs) provides a direct connection between the underlying neutrino-mass parameters and the phenomenology of RHNs at terrestrial experiments. We focus on $M_N\in[0.1,10]~\mathrm{GeV}$, a mass range in which beam-dump experiments and dedicated long-lived-particle searches offer particularly interesting opportunities to test small active–sterile mixing.

Fig.~\ref{fig:ship} presents the predicted parameter space in the $M_N - |\theta_e|^2$ plane, together with existing experimental constraints and the projected sensitivities of future facilities. The scatter points represent parameter space compatible with our analysis obtained by scanning over the Majorana phase $\alpha$ in the range $0<\alpha<2\pi$, and the relevant mass parameters: RHN mass $M_N$ from $0.1 - 10~\rm{GeV}$, and the ratio of the two Dirac mass parameters $m_1^D/m_2^D$ from $1$ to $10^{4}$ while $m^D_1\geq m^D_2$ is considered to avoid double counting. The colour coding, indicated in the right axis in $\log_{10}[m^D_1/m^D_2]$, thereby displays how the predicted active-sterile mixing is correlated with the hierarchy between the two Dirac-mass parameters. 
For a specific $M_N$, the lower boundary of the parameter space (scattered region) corresponds to the limiting state: $m_1^D=m_2^D$, whereas the upper boundary of it is determined by the requirement of reproducing the observed baryon asymmetry of the Universe. The viable region is therefore bounded by two distinct conditions: the imposed hierarchy between the Dirac-mass parameters and the requirement of successful leptogenesis. This provides a direct link between the structure of the predicted mixing parameter space and the underlying dynamics responsible for generating the baryon asymmetry.

The comparison with experimental sensitivities reveals an encouraging prospect for testing this scenario. In particular, the projected reaches of SHiP and MATHUSLA overlap with a substantial part of the predicted parameter space, making these experiments especially relevant for probing the RHNs responsible for the underlying leptogenesis dynamics. Their sensitivity to small mixing is important because RHNs in this regime can have relatively long lifetimes, potentially producing displaced decays that provide distinctive experimental signatures. Other proposed facilities, including CODEX-b \cite{CODEX-b:2019jve} and FCC-ee, offer complementary sensitivity in parts of the parameter space, while existing constraints from experiments such as NA62 and FASER further restrict the allowed region. The different sensitivity regions therefore serve as a complementary coverage.

\section{Majoron Phenomenology}
\label{majoron}

As stated earlier, the Majoron $\chi$, associated with the spontaneous breaking of $U(1)_L$, is massless in the absence of any explicit symmetry breaking \cite{Chikashige:1980qk,Chikashige:1980ui,Gelmini:1980re}. The higher-dimensional explicit $U(1)_L$-breaking operators such as $m^2 (\Phi^2+\Phi^{\star2})$ would yield a tiny mass for the Majoron. We briefly discuss its possible cosmological implications in our framework. We assume the radial degree of freedom of $\Phi$, denoted by $\phi$, to be sufficiently heavy to decouple from the thermal bath, with negligible SM Higgs-$\phi$ mixing.
In our previous work~\cite{King:2024idj}, we explored the connection between Majoron cosmology and high-scale resonant thermal leptogenesis, demonstrating that the Majoron can constitute viable freeze-in dark matter while satisfying existing indirect-detection and neutrino-experiment constraints (see also ref. \cite{Manna:2022gwn}). In that framework, the parameters responsible for Majoron production also induce the RHN mass splitting required for resonant leptogenesis, thereby forcing the RHN mass to be $M_N\sim\mathcal{O}(10^6\!-\!10^{13})~\mathrm{GeV}$. 

In contrast, the present framework involves GeV-scale RHNs with a tiny vacuum mass splitting (uncorrelated with Majoron as dark matter), motivated by low-scale Higgs-decay leptogenesis and its potential testability at experiments such as SHiP. Additionally, if we particularly focus on a regime of baryon asymmetry satisfaction by Higgs decay leptogenesis where the ARS mechanism is ineffective (with $T_{\rm osc} < T_{\rm sph}$), then it further pushes toward smaller $v_\phi\sim\mathcal{O}(100~\mathrm{GeV}-10~\mathrm{TeV})$ and a substantially higher cutoff, $\Lambda\sim\mathcal{O}(M_{\rm Pl})$. This is a consequence of the mass splitting $\Delta M^0 = 2v_{\phi}^2/\Lambda$ being required to be smaller in this case as seen from Figs. \ref{fig:DM0-YB1}, \ref{fig:DM0-YB2}.

\begin{figure}[tb!]
\centering\includegraphics[width=1\linewidth]{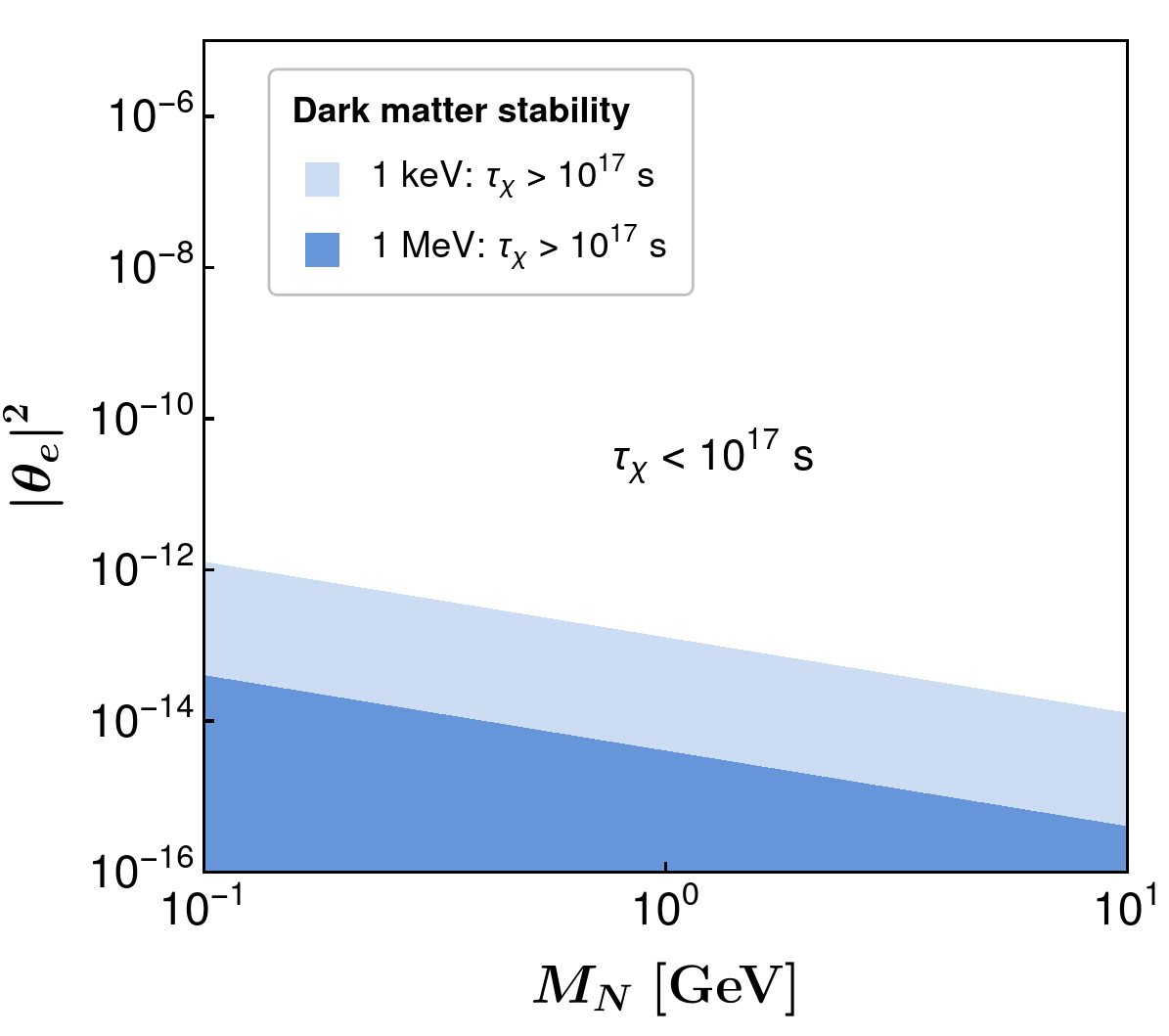}
	\caption{Viable regions in the $M_N-|\theta_e|^2$ plane for $v_\phi = 10^4~\mathrm{GeV}$, satisfying the Majoron lifetime constraint $\tau_\chi > 10^{17}\,\mathrm{s}$ for masses of $1~\mathrm{keV}$ (light blue) and $1~\mathrm{MeV}$ (dark blue).}
	\label{stability}
\end{figure}

On the other hand, Majoron can decay into light neutrinos through an effective interaction \cite{Escudero:2021rfi,Manna:2022gwn} $i\lambda_\nu\chi\bar{\nu}\gamma_5\nu$ originated from $\Phi-\mathcal{N}_1\mathcal{N}_2$ coupling of Eq. \eqref{eq:seesaw}, where $\lambda_\nu=m_\nu/v_\phi$. For $m_\chi\gg m_\nu$, the corresponding decay width is approximately
\begin{equation}
\Gamma_{\chi\to\nu\nu}\simeq
\frac{\lambda_\nu^2}{16\pi}m_\chi.
\end{equation}
Requiring the Majoron lifetime to exceed the age of the Universe, $\Gamma_{\chi\to\nu\nu}^{-1}>\tau_U$, therefore imposes a lower bound on $v_\phi$ that becomes increasingly stringent for larger $m_\chi$. As illustrated in Fig.~\ref{stability}, the blue-shaded region satisfies this stability condition, whereas the remaining region corresponds to a cosmologically unstable Majoron.

While successful baryogenesis favours smaller values of $v_{\phi}$, Majoron stability (in order to be dark matter) requires larger values, leading to a tension between these two requirements. Consequently, we find that the blue-shaded regions (in Fig. \ref{stability}) corresponding to the Majoron constituting the dark matter lies well below the scattered points in the $M_N-|\theta_{e}|^2$ plane. This indicates that, within the parameter space compatible with our low-scale leptogenesis scenario via Higgs decay, the Majoron cannot account for the observed dark matter abundance while simultaneously satisfying the stability requirement. Consequently, a keV - TeV Majoron is disfavoured as a candidate for stable cold dark matter in the parameter regime considered in this work, irrespective of its production mechanism.

Nevertheless, Majoron can still leave a cosmological imprint through its coupling into the neutrino sector. The relevant interaction ($i\lambda_\nu \chi \bar{\nu}\gamma_5\nu$) is precisely the generic neutrinophilic scalar coupling considered in the Ref. \cite{EscuderoAbenza:2020cmq}. As shown in that reference, such a neutrinophilic scalar, within the mass range ${\rm eV}<m_\chi<{\rm MeV}$ and with\footnote{A lower limit on $\lambda_{\nu}$ follows from 
$[v_{\phi}]_{\rm max}$ connected to the mass splitting using $\Lambda = M_{\rm Pl}$.\\} $\lambda_\nu<10^{-9}$, can enter into thermal equilibrium with the neutrino bath and later decay into the neutrinos after being non-relativistic, effectively reheating the neutrino bath relative to photons. This drives an increase in the effective number of relativistic species $N_{\rm eff}$.
Since the Majoron in our case shares the same regions of the parameters ($m_\chi,\lambda_\nu$), it offers a model-independent $\Delta N_{\rm eff}$ signature, testable by upcoming CMB experiments such as the Simons Observatory \cite{SimonsObservatory:2018koc} and CMB-S4 \cite{CMB-S4:2016ple}, that can possibly be correlated with the Higgs decay leptogenesis. Therefore, a detailed study of Majoron cosmology and its associated $\Delta N_{\rm eff}$ and neutrino-interaction signatures remains an interesting avenue for further exploration within our low-scale leptogenesis framework.

\section{Gravitational waves}
\label{sec:GW}
In the present scenario, a sheet like toplogical defects popularly known as DWs are formed after the $Z_2$ symmtery is spontaneously broken when $\Phi$ obtains a non-zero vev. To remove these DWs, one can introduce the following dimension-five non-renormalizable operators that explicitly break both the $Z_2$ symmetry and $U(1)_L$:
\begin{eqnarray}
	-\mathcal{L}_{\sla{Z_2}}&=& \frac{\beta_1}{\Lambda}(\Phi^\dagger \Phi)^2\Phi+  \frac{\beta_2}{2\Lambda}\left[\Phi^2+(\Phi^*)^2\right] \overline{\mathcal{N}_1^C}\mathcal{N}_2 \nonumber\\&+&\frac{\beta_3}{\Lambda}\overline{L_\alpha} \tilde{H}\mathcal{N}_1(\Phi+\Phi^*)+h.c.
\label{eq:dim5-Z2explicit}	
\end{eqnarray}
Introduction of these operators follow the argument that any global symmetry can be broken at the quantum gravity level \cite{Giddings:1988cx,Coleman:1988tj,Rey:1989mg,Abbott:1989jw,Barbieri:1979hc,Akhmedov:1992hh}.  Although the second and third terms in Eq.~\eqref{eq:dim5-Z2explicit} can, in principle, modify the RHN mass matrix and affect the generation of the lepton asymmetry, their contributions can be made negligible by considering $\beta_{2,3}\ll 1$. Moreover, such small values of $\beta_{2,3}$ are also preferred from the perspective of the gravitational-wave signals, as will become evident in the following discussion.


Once the DWs are formed, their energy density dilutes slower ($\rho_{\rm DW}\propto t^{-1}$) in comparison to that of radiation and matter, causing them to rapidly dominate the energy budget of the Universe. This famous DW problem can be resolved by destabilizing the walls through explicit breaking of the discrete symmetry \cite{Gelmini:1988sf} and introducing biased conditions at the time of formation. In this work, since we consider $Z_2$ symmetries are broken at dim-5, this allows us to write the following bias term (here we only provide the terms that will dominate the bias) with $\beta_i=\beta$:
\begin{equation}
 V_{\rm bias}= \beta\frac{v_\phi^5}{2\sqrt{2}\Lambda}. 
 \label{bias}
 \end{equation}
This allows the DW network to annihilate prior to dominating the Universe's energy budget, efficiently producing GWs in the process~\cite{Saikawa:2017hiv,Roshan:2024qnv,King:2023ayw,King:2023ztb,Bhattacharya:2023kws,Borah:2024kfn, Gouttenoire:2025ofv,Kitajima:2023cek,Vanvlasselaer:2026fay}. 

Assuming that the DW annihilate instantaneously ($t=t_{\text{ann}}$) during the radiation-dominated era, the peak frequency $f_{p}$ and peak energy density spectrum $\Omega_{p}h^2$ of GW at present can be expressed as \cite{Saikawa:2017hiv, Chen:2020wvu}
 \begin{widetext}
 \begin{align}\label{fpeak}
     f_{p}&\simeq 3.75\times10^{-9}~\text{Hz}\times C_{\rm ann}^{-1/2}\mathcal{A}^{-1/2}\bigg(\frac{10^3~\text{GeV}}{\mathcal{\sigma}^{1/3}}\bigg)^{3/2}\bigg(\frac{V_{\text{bias}}^{1/4}}{10^{-3}~\text{GeV}}\bigg)^{2}\,,\nonumber\\
     \Omega_{p}h^2&\simeq 5.3\times10^{-20}\times\tilde{\epsilon}_{\text{GW}}~C_{\rm ann}^{2}\mathcal{A}^{4}\bigg(\frac{\sigma^{1/3}}{10^3~\text{GeV}}\bigg)^{12}\bigg(\frac{10^{-3}~\text{GeV}}{V_{\text{bias}}^{1/4}}\bigg)^{8},
 \end{align}
 \end{widetext}
 where $\tilde{\epsilon}_{\text{GW}}\simeq0.7$~\cite{Hiramatsu:2013qaa} denotes the fraction of energy radiated into GW and can be regarded as a constant in the scaling regime. $\sigma=\sqrt{\frac{8\lambda_\phi}{9}}v_\phi^3$ denotes the surface tension of the DW. The GW spectrum from DWs characteristically exhibits a broken power-law form. The break frequency is set by the annihilation time, while the peak amplitude is determined by the DW energy density, as shown in Eq. \eqref{fpeak}. To describe this spectrum, we employ the broken power-law parametrization from Refs.~\cite{Caprini:2019egz, NANOGrav:2023hvm}:
\begin{eqnarray}
 \Omega_{\rm GW}h^2_{} = \Omega_p h^2 \frac{(a+b)^c}{\left(b x^{-a / c}+a x^{b / c}\right)^c} \ ,
\label{eq:spec-par}
\end{eqnarray}
where $x \equiv f/f_p$, and $a$, $b$ and $c$ are real and positive parameters. Here the low-frequency slope $a = 3$ can be fixed by causality, while numerical simulations suggest $b \simeq c \simeq 1$~\cite{Hiramatsu:2013qaa}.

\begin{figure}[tb!]
	\includegraphics[width=0.9\linewidth]{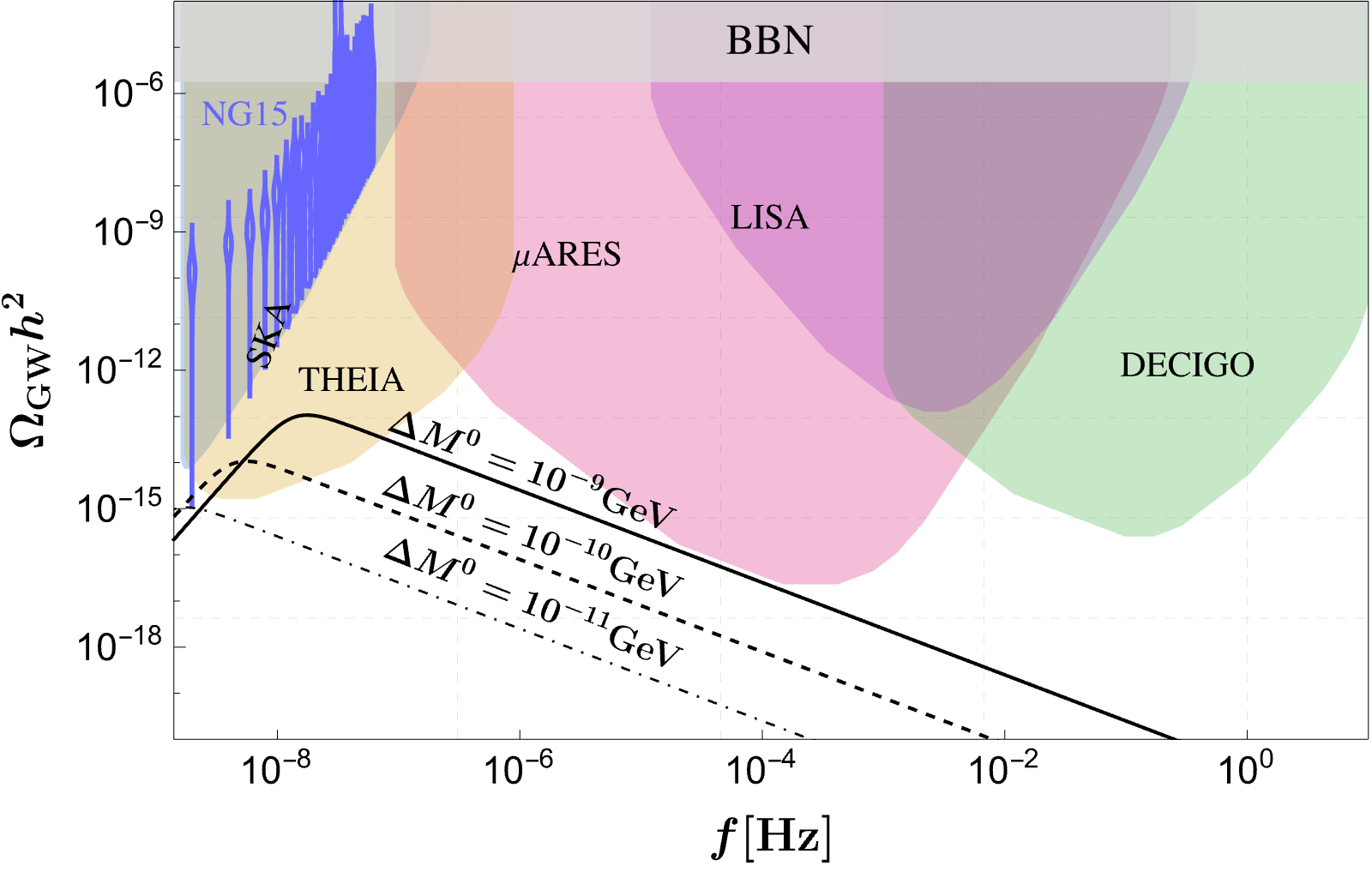}
	\caption{The GW spectrum produced from the annihilation of DWs for three different values of $\Delta M^0$. The plot is made for a fixed value of $\lambda_\phi=1$ , $\Lambda=M_{\rm Pl}$ and $\beta=10^{-10}$.}
	\label{fig:GW}
\end{figure}

In Fig. \ref{fig:GW}, we show the power-law integrated sensitivity curves \cite{Thrane:2013oya} of the future GW detectors ET~\cite{Punturo:2010zz}, LISA~\cite{LISA:2017pwj}, DECIGO~\cite{Kawamura:2020pcg}, $\mu$Ares~\cite{Sesana:2019vho}, SKA~\cite{Janssen:2014dka}, and THEIA~\cite{Garcia-Bellido:2021zgu}, the gray shaded region (labeled BBN) represents the constraint $ \Omega_{\rm GW}^{\rm \Delta N_{\rm eff}}h^2$ \cite{Domenech:2020ssp}. Next, we also show the GW curves in black (dashed and solid) resulting from the DW annihilation in Fig. \ref{fig:GW}. Following Eq. \eqref{fpeak}, it is clear that both the peak amplitude and peak frequency depends on four parameters: $\lambda_\phi,v_\phi$, $\Lambda$ and $\beta$. For the purpose of demonstration, we set $\lambda_\phi=1$. In order to connect the GW signals with leptogenesis parameter $\Delta M^0$, we rewrite $v_\phi$ as 
\begin{eqnarray}
    v_\phi=\sqrt{\frac{\Delta M^0 \Lambda}{2}}.
    \label{vphi_GW}
\end{eqnarray}
Substituting Eq. \eqref{vphi_GW} in Eq. \eqref{fpeak}, one finds $f_p\propto \sqrt{\beta\Delta M^0}$ and $\Omega_p h^2\propto \frac{\Lambda^3\Delta M^0}{\beta^2}$ for a fixed $\Lambda$ and $\beta$. While the peak frequency remains independent of $\Lambda$, the effect of $\Lambda$ can clearly be seen on the peak amplitude of the GW signal.
To obtain the black curves in Fig. \ref{fig:GW}, we set $\Lambda=M_{\rm Pl},~\beta=10^{-10}$ and choose three different values of $\Delta M^0$, namely, $\Delta M^0=10^{-9}$ GeV,  $10^{-10}$ GeV and $10^{-11}$ GeV  (consistent with leptogenesis). As expected, a larger $\Delta M^0$ corresponds to a larger peak frequency and larger peak amplitude of the GW signals.

\begin{figure}[tb!]
	\includegraphics[width=0.9\linewidth]{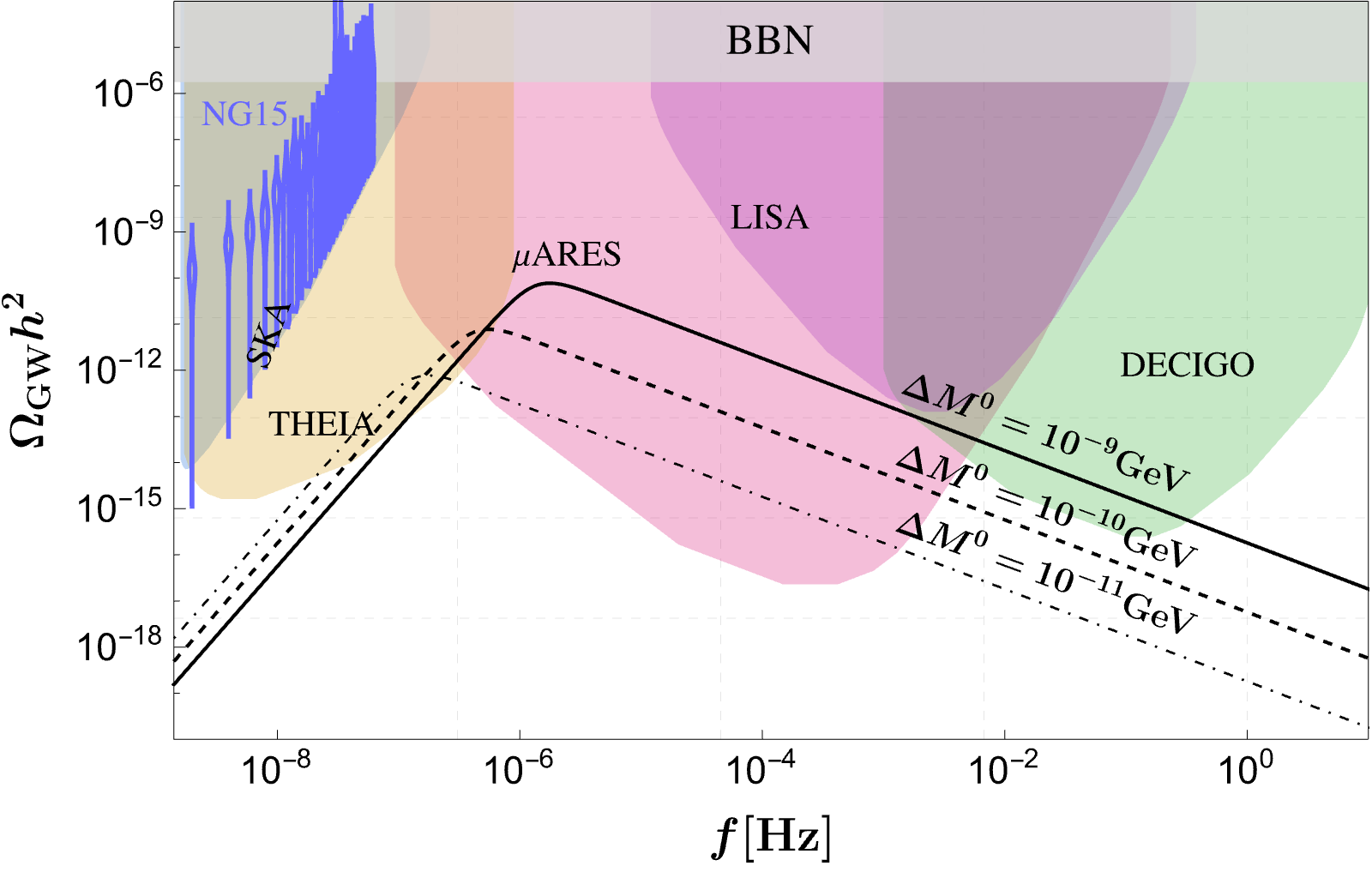}
	\caption{The GW spectrum produced from the annihilation of DWs for three different values of $\Delta M^0$. The plot is made for a fixed value of $\lambda_\phi=1$, $\Lambda^{\rm eff}=10^{22}$ GeV and $\beta=10^{-6}$.}
	\label{fig:GW2}
\end{figure}

Now we would also like to point out that the cutoff scale $\Lambda$ associated with higher-dimensional operators that break the global symmetry is typically taken to be the Planck mass $M_{\rm Pl}$ \cite{Addazi:2021xuf}. This choice is natural when symmetry breaking arises from perturbative gravitational effects. However, non-perturbative effects can substantially suppress the symmetry-breaking effects, yielding an effective cutoff scale that exceeds $M_{\rm Pl}$ by many orders of magnitude:
\begin{equation}
\Lambda^{\rm eff} = M_{\rm Pl} e^{\mathcal{S}} \gg M_{\rm Pl},
\end{equation}
where $\mathcal{S}$ denotes the instanton action. In this case the amplitude of GW signal would be enhanced. To demonstrate it, we set $\Lambda^{\rm eff}=10^{22}$ GeV while keeping the same three values of $\Delta M^0$ and setting $\beta=10^{-6}$ in Fig. \ref{fig:GW2}. As expected, increasing the value of $\Lambda^{\rm eff}$ increases the amplitude of GW signal.   
\noindent We realize that a tiny explicit breaking of the $Z_2$ symmetry can provide interesting GW signals that can be detected by the current and upcoming GW experiments providing an indirect probe of low-scale leptogenesis.
\section{Summary and Conclusion}
\label{conclusions}

In this work, motivated by the forthcoming experimental searches for heavy neutral leptons, we have discussed Higgs decay leptogenesis in a model with a quasi-Dirac neutrino
in the mass range 0.1-10 GeV, comprising two closely degenerate RHN states.
The model is based on the type-I seesaw mechanism with spontaneously broken $U(1)_L$ global lepton number, together with a discrete $Z_2$ symmetry, yielding two quasi-degenerate RHNs which form a single Dirac neutrino. 

We emphasise that the tiny mass splitting required between two quasi-degenerate RHNs arises naturally from the spontaneous and explicit breaking of the global $U(1)_L$ symmetry. In contrast to conventional resonant leptogenesis, where such a small mass splitting is typically introduced as an {\it ad hoc} manner, our construction provides a well-motivated symmetry origin for it through higher-dimensional operators. We have shown that the CP-asymmetry responsible for leptogenesis is entirely thermal in origin, arising from the interplay between the symmetry-induced vacuum mass splitting and finite-temperature corrections to the RHN masses. Solving the relevant Boltzmann equations, we have demonstrated that the observed baryon asymmetry can be successfully generated through thermal Higgs decays for RHN masses in the sub-electroweak regime.

We have identified the viable parameter space, consistent with Higgs decay leptogenesis, in terms of the quasi-Dirac neutrino mass 
range 0.1-10 GeV and active-sterile neutrino mixing, showing that a significant part of it falls within the projected sensitivities of future facilities such as SHiP, MATHUSLA and FCC-ee, thereby providing a realistic opportunity to test this scenario experimentally. We have also briefly discussed other viable realizations of low-scale leptogenesis, including the ARS mechanism based on RHN oscillations, and delineated the parameter space where Higgs decay leptogenesis dominates.

Finally we have shown that the symmetry-breaking sector of the model, in the relevant parameter range, gives rise to unstable domain walls whose annihilation can generate a stochastic gravitational-wave background, providing a complementary probe of the framework.

Taken together, these results establish a direct phenomenological connection between the conditions required for successful Higgs-decay-induced leptogenesis and the prospects for discovering the associated RHNs. The overlap between the theoretically viable region and the projected sensitivities of future experiments is particularly significant: it indicates that the mechanism can be tested through searches for active–sterile neutrino mixing, rather than remaining accessible only through its cosmological consequences. Although the precise coverage depends on the RHN mass and the detailed parameter choice, future long-lived-particle searches could probe a substantial fraction of the viable parameter space. Fig.~\ref{fig:ship} therefore identifies a promising experimental avenue for testing the RHN sector underlying our leptogenesis scenario.


	\section*{Acknowledgements}
	 SFK acknowledges financial support from the STFC Consolidated Grant ST/X000583/1.
    SFK was funded by a Leverhulme Trust Emeritus Fellowship Grant.

	\bibliographystyle{apsrev4-1}
	\bibliography{ref,refs}
\end{document}